\documentclass[letter,10pt]{article}

\usepackage[utf8]{inputenc}
\usepackage{authblk}
\usepackage{framed,multirow}

\usepackage{graphicx}	
\usepackage{subfig}
\usepackage{sidecap}    
\usepackage{url} 		
\usepackage{hyperref}

\usepackage[numbers,sort&compress]{natbib}

\usepackage{amssymb}	
\usepackage{amsmath}
\usepackage{latexsym}
\usepackage{mathtools}

\usepackage{verbatim}	

\usepackage{tikz}
\usetikzlibrary{shapes.geometric, arrows, positioning, calc}
\tikzstyle{block} = [rectangle, rounded corners, minimum width=1cm, minimum height=1cm, text centered, draw=black]
\tikzstyle{arrow} = [thick,->,>=stealth]

\usepackage[ruled,linesnumbered]{algorithm2e}	

\usepackage{booktabs}	
\usepackage{multirow}

\title{An Application of Manifold Learning in \\Global Shape Descriptors}

\author[1,4]{Fereshteh S. Bashiri$^{*,}$}
\author[2]{Reihaneh Rostami}
\author[4]{Peggy Peissig}
\author[3]{Roshan M. D'Souza}
\author[1,2]{Zeyun Yu\thanks{Corresponding authors.}$^{,}$}
\affil[1]{Department of Electrical Engineering, University of Wisconsin-Milwaukee, WI, USA}
\affil[2]{Department of Computer Science, University of Wisconsin-Milwaukee, WI, USA}
\affil[3]{Department of Mechanical Engineering, University of Wisconsin-Milwaukee, WI, USA}
\affil[4]{Center for Computational and Biomedical Informatics, Marshfield Clinic Research Institute, WI, USA}
\date{\today}

\providecommand{\keywords}[1]{\textbf{\textit{Keywords:}} #1}

\begin{document}

\maketitle

\begin{abstract}
    With the rapid expansion of applied 3D computational vision, shape descriptors have become increasingly important for a wide variety of applications and objects from molecules to planets.  Appropriate shape descriptors are critical for accurate (and efficient) shape retrieval and 3D model classification. Several spectral-based shape descriptors have been introduced by solving various physical equations over a 3D surface model. In this paper, for the first time, we incorporate a specific group of techniques in statistics and machine learning, known as \textit{manifold learning}, to develop a global shape descriptor in the computer graphics domain. The proposed descriptor utilizes the \textit{Laplacian Eigenmap} technique in which the Laplacian eigenvalue problem is discretized using an exponential weighting scheme. As a result, our descriptor eliminates the limitations tied to the existing spectral descriptors, namely dependency on triangular mesh representation and high intra-class quality of 3D models. We also present a straightforward normalization method to obtain a scale-invariant descriptor. The extensive experiments performed in this study show that the present contribution provides a highly discriminative and robust shape descriptor under the presence of a high level of noise, random scale variations, and low sampling rate, in addition to the known isometric-invariance property of the Laplace-Beltrami operator. The proposed method significantly outperforms state-of-the-art algorithms on several non-rigid shape retrieval benchmarks.
\end{abstract}


\section{Introduction}
\label{section:intro}
Three-dimensional models are ubiquitous data in the form of 3D surface meshes, point clouds, volumetric data, etc. in a wide variety of domains such as material and mechanical engineering~\cite{omrani16}, genetics~\cite{ng2007neuroinformatics}, molecular biology~\cite{gao2016mesh}, entomology~\cite{Sosa20163d7743319}, and  dentistry~\cite{riehemann2011microdisplay,wu2016model}, to name a few. Processing such large datasets (e.g., shape retrieval, matching, or recognition) is computationally expensive and memory intensive. For example, to query against a large database of 3D models to find the closest match for a 3D model of interest, one needs to develop an appropriate similarity measure as well as an efficient algorithm for search and retrieval. Shape descriptors assist with the example problem by providing discriminating feature vectors for shape retrieval~\cite{aflalo2011deformable,bronstein2011shape} and play a  fundamental role when dealing with shape analysis problems such as shape matching~\cite{xie2015deepshape,toldo2009visual} and classification~\cite{bu2014learning}. 

In general, there are two types of shape descriptors: \textit{local descriptors}, also called \textit{point signatures}, and \textit{global descriptors}, referred to as \textit{shape fingerprints}. A local shape descriptor computes a feature vector for every point of a 3D model. On the other hand, a global shape descriptor represents the whole 3D shape model in the form of a low-dimension vector. A descriptor that is informative and concise captures as much information as possible from the 3D shape including the geometric and topological features. Such a vector drastically lowers the shape analysis burdens in terms of both computational intensity and memory. 

While a large number of successful non-spectral shape descriptors have been proposed in the literature, spectral descriptors have proved to be beneficial in many applications~\cite{boscaini2016anisotropic,bronstein2011spectral}. The spectral methods take advantage of eigen-decomposition of the Laplace-Beltrami (LB) operator applied on the shapes and construct their informative descriptors using the eigenvalues and eigenvectors. These methods have found successful applications in graph processing~\cite{raviv2013graph}, computational biology~\cite{de2013isometry}, and point-to-point correspondence~\cite{ovsjanikov2012functional}. 

One of the first spectral descriptors introduced to the computer graphics community is Shape-DNA, developed by Reuter et al. in 2006~\cite{reuter2006laplace}. Shape-DNA attracted a great deal of attention for its unique isometric and rotation invariant features~\cite{reuter2006laplace}. Since then, several local as well as global shape descriptors have been introduced in accordance with Shape-DNA such as Heat Kernel Signature (HKS)~\cite{sun2009concise}, Wave Kernel Signature (WKS)~\cite{aubry2011wave}, and Global Point Signature (GPS)~\cite{rustamov2007laplace}. The common ground between these methods is the discretization approach used to solve the Laplacian eigenvalue problem, which uses a cotangent weighting scheme along with area normalization.

Although there are many advantages of using variations of the cotangent scheme, there are several limitations. First, by their nature, they are limited to triangulated meshes. Second, they do not perform well when dealing with degenerate and non-uniform sampled meshes~\cite{reuter2009discrete,belkin2008discrete}. Also, their convergence error depends on factors such as the linearity of the function on the surface~\cite{belkin2008discrete}. One possible approach to address these limitations is through the use of manifold learning.



Nonlinear dimensionality reduction techniques, known as \textit{manifold learning}, assume the existence of a low-dimensional space, which represents a high-dimensional manifold without much loss of information \cite{goldberg2008manifold}. Similar to global descriptors, manifold learning methods attempt to learn the geometry of a manifold in order to extract a low dimensional vector of features that is informative and discriminative. However, unlike shape descriptors, the number of dimensions of a space does not confine manifold learning methods. To the best of our knowledge, the application of manifold learning, an active research topic in statistics and machine learning, has not been investigated in the computer graphics community for extracting global shape descriptors. This motivates the primary aim of this research, which is to explore the effectiveness of a manifold learning method, more specifically \textit{Laplacian Eigenmap} \cite{belkin2003laplacian}, in representing a 3D model with a low-dimensional vector. Our work introduces a novel Laplacian Eigenmap-based global shape descriptor and provides a straightforward normalization method that significantly outperforms existing state-of-the-art approaches.

 
In our \textbf{first contribution}, inspired by the idea of Laplacian Eigenmaps~\cite{belkin2003laplacian}, we learn the manifold of a 3D model and then, analogous to the approach taken by Shape-DNA, we make use of the spectrum of the embedded manifold to build the global shape descriptor. This approach has two main advantages. First, it relies on the adjacency of the nodes, disregarding the fine details of the mesh structure. Therefore, it can be used for degenerate or non-uniform sampled meshes. Second, as manifold learning does not rely on the mesh structure and is not limited to a specific type of meshes, e.g., triangulated meshes, it can be applied easily to any other mesh types such as quadrilateral meshes.


In our \textbf{second contribution}, we present a simple and straightforward normalization technique (motivated by~\cite{reuter2006laplace,bronstein2010scale,kuang2015modal}) to obtain a scale-invariant global shape descriptor that is more robust to noise. To this end, we propose to subtract the first non-zero eigenvalue from the shape descriptor after taking the logarithm of the spectrum. One advantage of our approach over the idea of Bronstein et al.~\cite{bronstein2010scale} is that we avoid taking the direct derivative; this advantage is significant since the differential operator amplifies the noise. Taking the logarithm additionally helps to suppress the effect of the noise that is present in higher order elements of the spectrum.

The remainder of this paper is organized as follows. In Section~\ref{sec:background}, we briefly overview spectral shape analysis and manifold learning. Then in Section~\ref{section:methods}, we introduce the proposed shape descriptor along with some technical background. In Section~\ref{sec:experiments}, the performance of the proposed method, as well as the robustness of the algorithm are examined and compared with multiple well-known shape descriptors by performing several qualitative and quantitative experiments using widely used 3D model datasets. Section~\ref{sec:discuss} discusses the results in more detail and draws conclusions.

\section{Background}
\label{sec:background}
In this section, we first review spectral shape analysis, more specifically global shape fingerprints, and different discretization methods of the LB operator. Then, we briefly review manifold learning, more specifically Laplacian Eigenmap, to provide the necessary foundation for developing our proposed Laplacian Eigenmap based scale-invariant shape descriptor, which from now on we call LESI. 

\subsection{Spectral Shape Analysis}
\label{subsec:spectral}
The \textit{Laplace-Beltrami operator} $\Delta$ is a linear differential operator defined on the differentiable manifold $\mathcal{M}$ as the divergence of the gradient of a function $f$ as the following form~\cite{reuter2006laplace,gao2014compact}:
\begin{equation}
\Delta f = div(grad(f)).
\end{equation} 

L\'{e}vy \cite{levy2006laplace} noted that the eigenfunctions $\phi_i$ of the continuous LB operator, which are the solution to the following \textit{Laplacian eigenvalue problem}:
\begin{equation}
\label{eq:Laplacian Eigenvalue}
\Delta f = -\lambda f
\end{equation}
are \textit{the orthogonal} basis for the space of functions defined on the surface of a manifold. In other words, a function $f$ on the surface can be expressed as a sum over coefficients of these infinite bases:
\begin{equation*}
f = c_0 \phi_0 + c_1 \phi_1 + ...
\end{equation*}

Furthermore, the LB operator is positive semi-definite, having non-negative eigenvalues $\lambda_i$ that can be sorted as follows:
\begin{equation*}
0 \le \lambda_1 \le \lambda_2 \le ... \le \lambda_i \le ...
\end{equation*}
The sequence of eigenvalues of the LB operator is called the \textit{spectrum} of the LB operator. As it is computed based on the gradient and divergence that depend on the Riemannian structure of the manifold, it possesses the isometry invariant property~\cite{reuter2006laplace}.

These significant features of the LB operator, which include the orthogonal basis and non-negative spectrum, motivated researchers to develop various local and global shape descriptors. The Shape-DNA and HKS were developed by considering the heat distribution as the function $f$ on the surface. The WKS was obtained by solving the Schr\"{o}dinger wave equation on the surface of the manifold. Also, it has been shown that the GPS descriptor is in close relation to the Green's function on the surface~\cite{rustamov2007laplace}. 

To approximate equation \eqref{eq:Laplacian Eigenvalue}, despite the choice of function $f$, we need a discretization scheme. Different discretization schemas (e.g., Taubin~\cite{taubin1995signal}, Mayer~\cite{mayer2001numerical}) of the LB operator on the triangular meshes are discussed in~\cite{xu2004discrete}. 

A 3D shape, sampled from the surface of a Riemannian manifold $\mathcal{M}$, is usually presented by a set of vertices $V$ and their connectivity $E$ in the form of the graph $G=(V,E)$. For a surface mesh $G$, according to \cite{reuter2009discrete}, the equation \eqref{eq:Laplacian Eigenvalue} can be discretized as:
\begin{equation}
A \textbf{f} = \lambda B \textbf{f}
\end{equation}
where $A$ is the \textit{stiffness matrix} and $B$ is the \textit{lumped mass matrix}. One popular approach to constructing the matrix $A$, is using weights:
\begin{equation}
w_{ij} = \frac{\cot(\alpha_{ij})+\cot(\beta_{ij})}{2}
\end{equation}
where $\alpha_{ij}$ and $\beta_{ij}$ are the two angles facing the edge $(i,j)$. Different mass normalization methods using the triangle area or the Voronoi region area are suggested to construct the matrix $B$. The cotangent weighting schema and its variants have been utilized in multiple FEM-based discretization methods. 

Another approach to constructing the matrix $A$ is to use the heat kernel weight, also known as the exponential weighting scheme, as follows:
\begin{equation}
w_{ij} = e^{- \frac{{\|x_i-x_j\|}^2}{t}}
\end{equation}
where $\|.\|$ denotes the Euclidean distance between two adjacent nodes $i$ and $j$. 

In \cite{reuter2009discrete}, several existing discretization methods, including variants of linear FEM \cite{desbrun1999implicit,meyer2003discrete} and heat kernel weighting proposed in~\cite{belkin2008discrete} are compared. According to \cite{reuter2009discrete} and from the discussion led by Xu in \cite{xu2004discrete,xu2006convergence}, discrete LB operator using cotangent weighting scheme may not converge in all situations, specifically when dealing with non-uniform meshes. However, the heat kernel weighting scheme proposed in \cite{belkin2008discrete} does not depend on the peculiarities of the triangulation and outperforms all linear approaches \cite{reuter2009discrete,sun2009concise}. In addition, concerning the type of the function $f$, the cotangent scheme only converges for linear functions, while the heat kernel scheme converges well for nonlinear functions as well \cite{belkin2008discrete}. The proposed exponential approximation scheme provides point-wise convergence with good stability with respect to noise. It is important to note that although the method was discussed for surfaces without boundary, the results are valid for interior points of a surface with boundary~\cite{belkin2008discrete}.

\subsection{Manifold Learning}
\label{subsec:manifold}

To make the current contribution self-contained, we provide a brief introduction from the data analysis perspective. Dimensionality reduction of high-dimensional data is a critical step in data analysis and processing. Non-linear dimensionality reduction, also known as \textit{manifold learning}, is a problem of finding a low-dimensional representation for high-dimensional data. Several local and global manifold learning methods have been developed including Isomap~\cite{tenenbaum2000global, silva2003global}, LLE~\cite{roweis2000nonlinear, saul2003think}, Laplacian Eigenmap~\cite{belkin2003laplacian}, and Diffusion maps~\cite{coifman2006diffusion}. We review these next.

Consider a set $x_1,...,x_n\in \mathcal{M}$ of $n$ points on manifold embedded in $\mathbb{R}^l$. Manifold learning methods look for a set of corresponding points $y_1,...,y_n$ in $\mathbb{R}^m (m\ll l)$ as a structural representation, while respecting some local or global information. Each method attempts to minimize a cost function in this mapping.

Laplacian Eigenmap \cite{belkin2002laplacian}, proposed by Belkin and Niyogy in 2002, is a computationally efficient and mathematically well-studied manifold learning technique. It is based upon graph Laplacian and Laplace-Beltrami operator on the manifold. Accordingly, it is considered as a spectral analysis method. Laplacian Eigenmap deals with sparse, symmetric, and positive semi-definite matrices. It is in close connection to the heat flow~\cite{belkin2002laplacian,belkin2003laplacian}.

Briefly speaking, for a given manifold, Laplacian Eigenmap applies the graph Laplacian operator and uses the eigenfunctions of such operator to provide the optimal embedding. Laplacian Eigenmap preserves local information by minimizing the distance between embedded points, which are mapped from adjacent points in the original high-dimensional space~\cite{belkin2003laplacian}. Aside from the locality preserving property, it provides structural equivalence and discrimination by capturing the intrinsic geometric structure of the manifold. The structural equivalence property states that two similar manifolds will have similar representation after projecting into a lower dimension space~\cite{wachinger2010manifold,wachinger2010structural}.

Some other manifold learning methods, e.g., Isomap, LLE, and Diffusion map are also based on spectral analysis of the high-dimensional manifold. In contrast to these methods that construct the orthogonal basis of their desired low-dimensional space using eigenfunctions of an LB operator, we develop our scale-invariant shape descriptor using the spectrum of the LB operator. Even though our primary focus is on the Laplacian Eigenmap, owing to its unique properties, we believe that other spectral manifold learning methods are also capable of extracting informative and discriminative shape fingerprints.

\section{Method}
\label{section:methods}
\begin{figure*}[ht!]
\begin{center}
\begin{tikzpicture}[node distance=3.5cm, text width=2.8cm]
\scriptsize
\node (lmat) [block] {Construct: \\ Laplacian matrix ($L$)};
\node (eigp) [block, right of=lmat] {Solve: \\ $Lf=\lambda D f$};
\node (eval) [block, below of=eigp, node distance=1.5cm] {Spectrum: \\ $0<\lambda_1 \le... \le \lambda_m $};
\node (norm) [block, left of=eval] {Normalization: \\ $log(\lambda_i)-log(\lambda_1)$ \\$ 1\le i\le m$};

\draw [arrow] ++(-2,0) node [left, text width=1cm] {3D model} -- (lmat.west);
\draw [arrow] (lmat) -- (eigp);
\draw [arrow] (eigp) -- (eval);
\draw [arrow] (eval) -- (norm);
\end{tikzpicture}

\caption{The block diagram of the proposed Laplacian Eigenmaps based scale-invariant (LESI) global shape descriptor.}
\label{fig:Workflow}
\end{center}
\end{figure*}
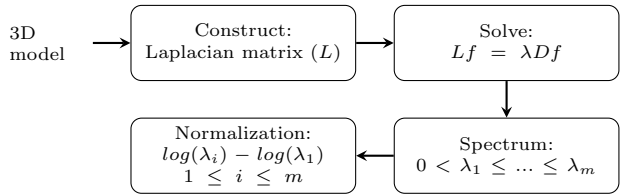

In this section, we elaborate our proposed LESI global shape descriptor. A flowchart of the proposed approach is shown in Figure~\ref{fig:Workflow}.

\subsection{Laplacian Eigenmap-Based Shape Descriptor}
\label{subsec:le descriptor}
We treat a global descriptor as a dimensionality reduction problem as it squeezes the latent information of a 3D model into a vector. Due to the fact that the Laplacian Eigenmap has two properties of structural equivalence and locality preservation~\cite{belkin2003laplacian}, we propose a global shape descriptor using the spectrum of graph Laplacian. 

A graph Laplacian is constructed over an \textit{undirected weighted graph} $G=(V,E)$ with a set of points $x_i\in V$ and a set $E$ of edges that connects nearby points ($(i,j)\in E$). The theory behind finding the optimal embedding in a Laplacian Eigenmap requires an undirected graph. Every 3D model is given in bidirectional connection and hence, we need to neither examine nor force it to the graph. However, as we will explain later, we need to remove isolated points. Considering the advantages of the heat kernel weighting scheme, which are summarized in Section \ref{subsec:spectral} and discussed in details in \cite{belkin2008discrete}, Laplacian Eigenmap suggests constructing the weighted graph as follows:

\begin{equation}
\label{eq:LE weight}
w_{ij}=
\begin{cases}
\begin{aligned}
&e^{- \frac{{\|x_i-x_j\|}^2}{t}} &\qquad& \textrm{, if } (i,j)\in E\\
&0  &&\textmd{, otherwise}
\end{aligned}
\end{cases}
\end{equation}

The only parameter in equation \eqref{eq:LE weight} is $t$, which defines the extent to which distant neighbors influence the embedding of each point. The choice of parameter $t$ is data-dependent, and there exists no unique way in the literature to select the proper value, but it can be tuned empirically. As $t$ has neither a very high impact on the final embedding nor the convergence rate of our final derivations, we empirically recommend:
\begin{equation}
\label{eq:scale t}
\begin{split}
t&=2 d^2_{max} \textrm{ where,} \\
d_{max}&=\max{\|x_i-x_j\|} \quad, \forall (i,j)\in E.
\end{split}
\end{equation}
Here, weights are bounded as $0.6 \approx e^{-0.5} \le w_{ij} \le 1$.

Laplacian Eigenmap attempts to find a low dimensional data set that preserves local information. For this purpose, it assumes two neighboring points $x_i$ and $x_j$ stay close after being mapped to $y_i$ and $y_j$. Therefore, it minimizes the following function~\cite{belkin2003laplacian}:
\begin{equation}
\label{eq:LE optimization}
\frac{1}{2}\sum\limits_{ij} (y_i-y_j)^2W_{ij}=\textbf{y}^TL\textbf{y}.
\end{equation}
where $L=D-W$ is the so called \textit{Laplacian matrix}, $W_{ij}$ is a symmetric \textit{weight matrix}, and $D_{ii}=\sum_j W_{ij}$, the \textit{degree matrix}, is a diagonal matrix. The assumption that the graph is undirected yields the symmetric property of $W$ and consequently, $D$ and $L$. It plays a critical role in deriving equation \eqref{eq:LE optimization}.

By adding the orthogonality constraint $\textbf{y}^TD\textbf{1}=0$ in order to eliminate the trivial solution and the constraint $\textbf{y}^TD\textbf{y}=1$ for removing an arbitrary scaling factor in the embedding, the minimization problem (\ref{eq:LE optimization}) simplifies to:
\begin{equation}
\label{eq:LE argmin}
\underset{\substack{\textbf{y}\\ \textbf{y}^TD\textbf{y}=1 \\ \textbf{y}^TD\textbf{1}=0}}{\arg\min} \textrm{ } \textbf{y}^TL\textbf{y}.
\end{equation}

The matrix $L$ is real, symmetric, and positive semi-definite. Therefore, the solution vector $\textbf{y}$ (in equation \eqref{eq:LE argmin}) is obtained by the minimum eigenvalue solution to the generalized eigenvalue problem~\cite{belkin2003laplacian}:
\begin{equation}
\label{eq:LE equation}
L\textbf{y}=\lambda D\textbf{y}.
\end{equation}

At this point, the optimal low dimensional embedding, suggested by the Laplacian Eigenmap, is obtained by utilizing the eigenvectors. However, we focus on the spectrum of the graph Laplacian and its' properties. Eigenvalues obtained from equation \eqref{eq:LE equation} are real, non-negative, and sorted in increasing order as follows:
\begin{equation*}
0 \le \lambda_1 \le \lambda_2 \le ... \le \lambda_n.
\end{equation*}

As the row (or column) sum of $L$ is equal to zero, eigenvalue $\lambda=0$ and a corresponding eigenvector \textbf{1} are trivial solutions to equation \eqref{eq:LE equation}. The multiplicity of eigenvalue zero is associated with the number of connected components of the graph. Eigen-solvers often obtain very small, though not precisely zero, eigenvalues due to the computational approximations. If we may know the number of connected components of $L$, we can discard all eigenvalues equal to zero, and form our shape fingerprint using the more informative section of the spectrum. This is easily done by Dulmage-Mendelsohn decomposition~\cite{dulmage1958coverings}.

The second smallest eigenvalue, also known as the \textit{Fiedler value}, is a measure of the connectivity within the graph. If the graph has $c$ connected components, our proposed shape descriptor is a set of $d$ eigenvalues as:
\begin{equation}
\textmd{LESI} \coloneqq (\lambda_{c+1}, \lambda_{c+2},\ldots,\lambda_{c+d})
\end{equation}

The LESI descriptor is composed of the spectrum of the LB operator, and hence, it is isometric invariant, independent of the shape location, and informative. The latter, discussed in spectral graph theory, states that the spectrum of graph Laplacian contains a considerable amount of geometrical and topological information of the graph. Moreover, LESI has the similarity property, caused by the structural equivalence property of Laplacian Eigenmap, meaning that two 3D models from the same class of models have similar fingerprints. Unlike Shape-DNA and other shape descriptors that are based on the cotangent weighting scheme, LESI is not limited to triangulated mesh structures because the Laplacian Eigenmap is capable of dealing with high-dimensional data. For some applications in which scale is not a determinant factor, it is favorable to have a scale-invariant descriptor. A fast and efficient normalization method is presented in Section~\ref{subsec:normalization}.

One important matter to consider is the convergence and accuracy of the proposed fingerprint, which ultimately depends on the heat kernel-based discretization of the LB operator. The cotangent weighting scheme and its variants are sensitive to the peculiarities and quality of the particular triangulation of the mesh (refer to Section \ref{subsec:spectral}). While an exponential weighting scheme has shown accurate performance in dealing with nonlinear functions over the surface and non-uniform mesh representations, it is not clear how this method can handle manifolds with boundaries~\cite{reuter2009discrete,belkin2008discrete}. It does, however, behave well for interior points of the surface. Therefore, we recommend removing rows and columns of $L$ and $D$ corresponding to isolated points, before solving equation \eqref{eq:LE equation}. The descriptor obtained from the rest of the connected graph is an informative and discriminative descriptor of the graph.

\subsection{Scale Normalization} 
\label{subsec:normalization}
For some applications, the size of an object is not a determinant factor in shape comparison and identification. Therefore, a scale-invariant shape descriptor with a solid normalization method is more desirable. For that purpose, some shape descriptors including Shape-DNA, have proposed multiple normalization methods. Most normalization methods of Shape-DNA focus on finding an appropriate scaling factor, such as the surface area, the volume, or coefficient of a fitting curve, which will be multiplied in the descriptor.

Moreover, it is shown that eigenvalues with a higher order are more susceptible to noise. For that reason, the original Shape-DNA recommends cropping the spectrum and using no more than 100 eigenvalues~\cite{reuter2006laplace}. In this section, we propose a simple and efficient normalization method that significantly reduces the effect of scale variations as well as noise. In this approach, we are interested in taking the scaling factor out from the descriptor in one step, rather than an extra step to find an appropriate neutralizing factor. Although the normalization seems simple, later in the experiments section, we will show its efficiency. The work presented in~\cite{sun2009concise} and \cite{reuter2006laplace} influenced this method. 

\begin{figure*}[t!]
    \captionsetup[subfigure]{justification=centering,width=0.2\textwidth}
    \centering
    \subfloat[Teddy Bear models\label{subfig:teddymodel}]{%
    \includegraphics[width=0.1\textwidth]{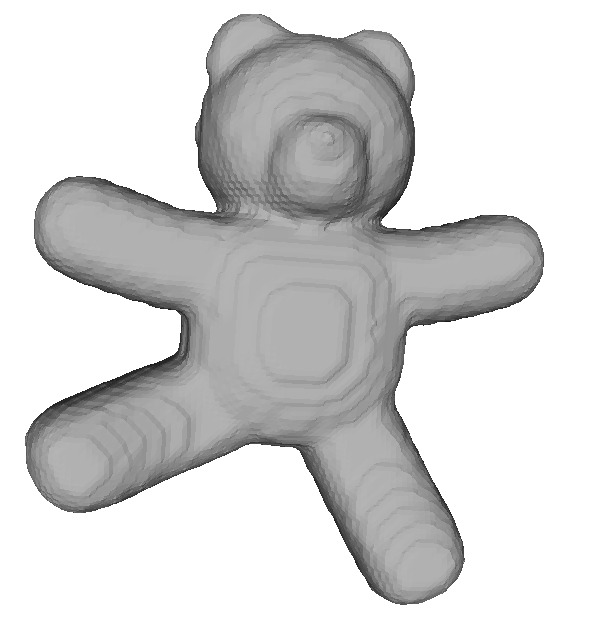}
    \includegraphics[width=0.1\textwidth]{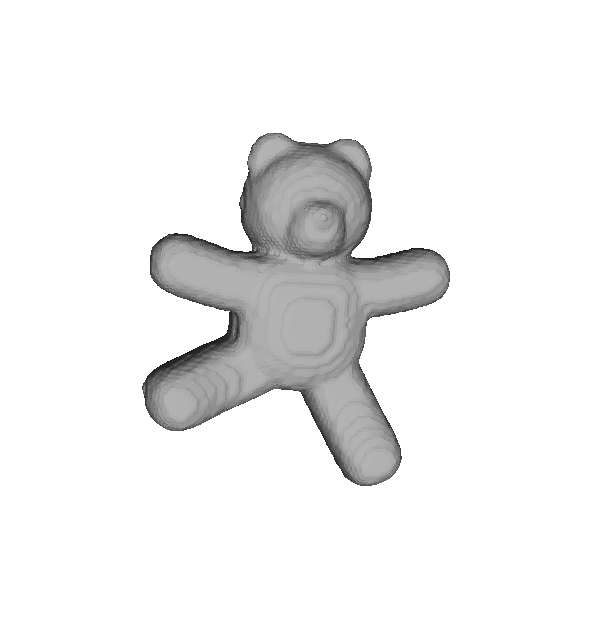}
    }
    \subfloat[Spectrum of LB operator\label{subfig:spectrum}]{%
    \includegraphics[width=0.25\textwidth]{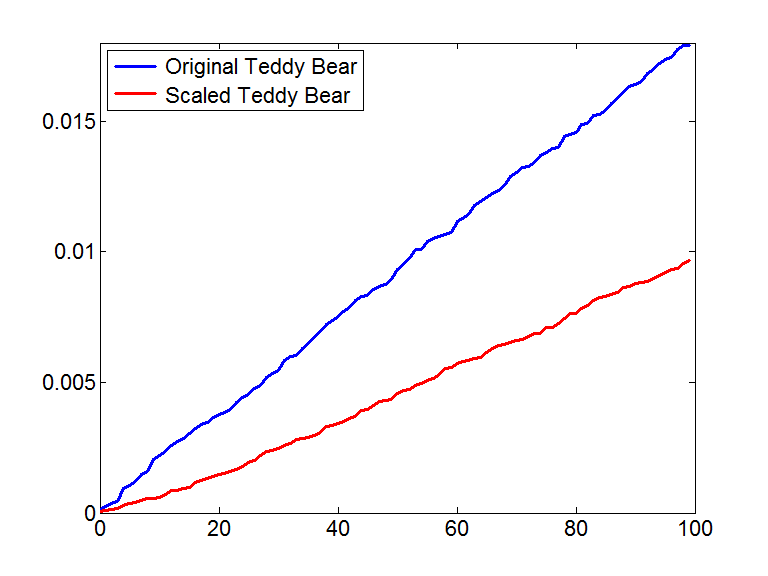}
    }
    \subfloat[Logarithm of spectrum\label{subfig:logspectrum}]{%
    \includegraphics[width=0.25\textwidth]{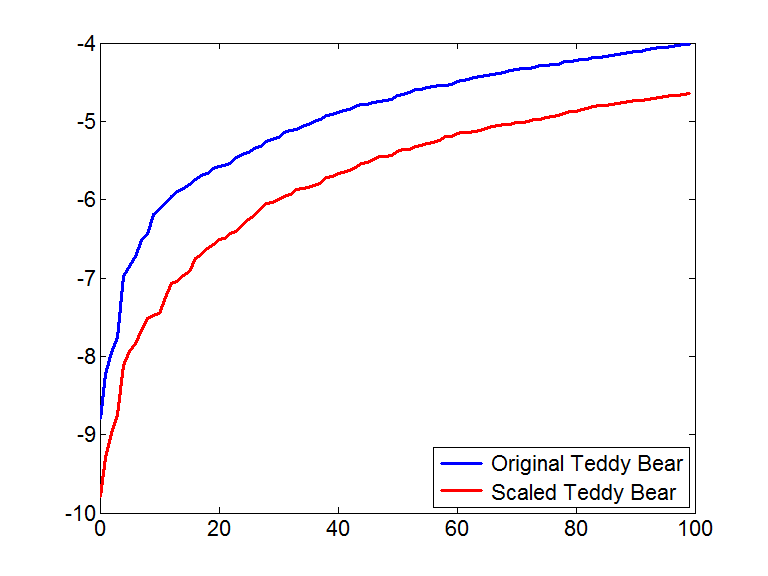}
    }
    \subfloat[Normalized Spectrum\label{subfig:normspectrum}]{%
    \includegraphics[width=0.25\textwidth]{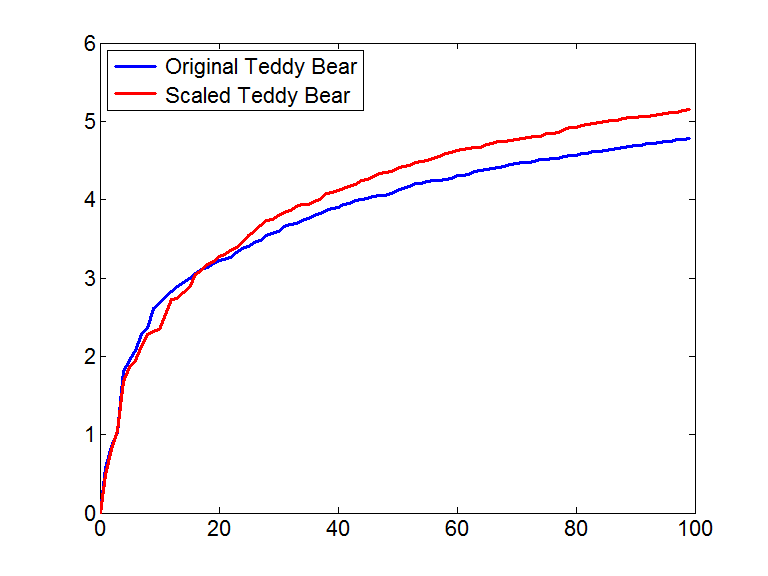}
    }\\
    \caption{An example showing the proposed normalization method of the shape descriptor. \protect\subref{subfig:teddymodel} The Teddy Bear model and its scaled version (scale factor 0.7). \protect\subref{subfig:spectrum} The spectrum of original (blue) and scaled (red) Teddy Bear models. Please note that the original spectrum is approximately multiplied by half. \protect\subref{subfig:logspectrum} The Logarithm of the spectrum shown in \protect\subref{subfig:spectrum}. \protect\subref{subfig:normspectrum} The normalized spectrum of original and scaled Teddy Bear models after subtracting first element of logarithm of the spectrum .}
	\label{fig:scaleinvariantexmple}
\end{figure*}

According to Weyl's law~\cite{weyl1911asymptotische}, when an arbitrary scale object is scaled with factor $\alpha$, the eigenvalues are scaled with factor $\alpha^2$. Let $\Lambda_s=(\alpha^2\lambda_1,\ldots,\alpha^2\lambda_d)$ be a LESI descriptor obtained from a scaled 3D model with unknown factor $\alpha$. To normalize the descriptor and eliminate the effect of the scaling factor, we recommend computing:
\begin{equation}
\Lambda_n(i)= \log(\Lambda_s(i)) - \log(\Lambda_s(1)) \textrm{ , for } 1\le i\le d
\end{equation}

To achieve a scale invariant shape descriptor, we first take the logarithm of the descriptor vector and then compute the difference of the new vector from its smallest element. This is equivalent to dividing the vector by its first element and taking the logarithm next. Basically, division takes the factor out, and the logarithm eliminates the influence of noise. 

Figure \ref{fig:scaleinvariantexmple} illustrates the details of the proposed algorithm in an example. In Figure~\ref{fig:scaleinvariantexmple}\subref{subfig:teddymodel}, two Teddy Bear models are shown. One model is in the original size whereas the other model is scaled with factor 0.7. It is clear from Figure \ref{fig:scaleinvariantexmple}\subref{subfig:spectrum} that the spectrum of the scaled model is almost $(0.7)^2$ of the spectrum of the original model. Taking the logarithm of the spectrum takes away the scaling factor from multiplicand and leaves it as augend. Therefore, subtracting one term (e.g., the first element) removes the scaling factor from all other terms. The result is a normalized and scale-free spectrum.


\subsection{Algorithm}
\label{subsec:algorithm}
Our proposed descriptor consists of three major steps. For a given 3D polygonal model $G=(V,E)$ with a set of vertices $V=(x_1,\ldots,x_n)\in \mathbb{R}^{3\times n}$ and a set of neighbor connections $E$, the LESI descriptor is a $d$-dimensional vector $\Lambda=(\lambda_1,\ldots,\lambda_d)$ of real and positive values.

In the first step, we compute the $n\times n$ real, symmetric, and sparse weight matrix $(W)$ for a 1-ring neighbor of every point as stated in equation \eqref{eq:LE weight} using the inner scaling factor given in equation \eqref{eq:scale t}. Next, we form the generalized eigenvalue problem equation \eqref{eq:LE equation} by constructing Laplacian and degree matrices ($L$ and $D$ respectively) without difficulty. The matrix $L$ is sparse, real, symmetric and semi-positive. Utilizing the Dulmage-Mendelsohn decomposition, we find the number of connected components of $L$. The objective of the second step is to find the spectrum of the LB operator. For that purpose, we solve the generalized eigenvalue problem using the Lanczos method. Then, we leave out as many smallest eigenvalues as the number of connected components. Since in most cases a single 3D model is made up of one connected component, we only need to leave out one eigenvalue. The last step of the algorithm deals with scale normalization and noise reduction, in case it is required, by taking the logarithm of the spectrum and subtracting the first element from the rest of the vector. Detailed steps of the algorithm are summarized in Algorithm \ref{algorithm}.

\begin{algorithm}[ht]
\SetAlgoLined
\SetKwInput{KwData}{Input}
\SetKwInput{KwResult}{Output}
\KwData{A 3D polygonal model $G=(V,E)$ with n vertices $x_i$ and edge list $E$}
\KwResult{A $d$-dimensional vector $\Lambda=(\lambda_1,\ldots,\lambda_d)$ }
Compute edge weights using equations \eqref{eq:LE weight} and \eqref{eq:scale t}\;
Construct the sparse, real, and symmetric $n \times n$ matrices $W$, $D$, and $L$\;
Find number of connected components (nConComp) from $L$\;
Solve equation \eqref{eq:LE equation} for $nConComp+d$ eigenvalues\;
Sort them in increasing order and leave out $nConComp$ smallest ones\;
\If{normalization is required}{
	Compute $log(\lambda_i)-log(\lambda_1)$ where \\$ 1\le i\le d$\;
   	}
\caption{Laplacian Eigenmap-based scale-invariant global shape descriptor}
\label{algorithm}
\end{algorithm}

\section{Experiments}
\label{sec:experiments}

In Section \ref{subsec:dataset}, we first present two datasets used in our experiments. Then, in Section~\ref{subsec:retrievalresuls}, we qualitatively visualize and measure the competence of the proposed method in discriminating different clusters compared with candidate methods from the literature. Next, in Section \ref{subsec:classification}, we validate the effectiveness of the LESI descriptor to distinct multiple classes by measuring the accuracy of multi-class classification. Finally, in Section~\ref{subsec:robustness}, extensive experiments are carried out to study the robustness of the proposed shape descriptor with respect to noise, scale invariance, and down-sampling. All the algorithms were implemented using the MATLAB R2013a environment running on a personal computer with Intel(R) Core(TM) i3-4130 CPU @ 3.40GHz and 12GB memory.

\subsection{Dataset}
\label{subsec:dataset}
To validate the utility of the proposed shape descriptor, we utilized two standard, widely-used, and publicly available datasets of 3D polygon meshes. The high-resolution TOSCA dataset \cite{bronstein2008numerical} contains 80 three-dimensional non-rigid models, including 11 cats, 6 centaurs, 9 dogs, 4 gorillas, 8 horses, 12 women poses, 3 wolves and two men with 7 and 20 poses respectively. In our experiments, we use all models except the gorilla models, as they contain isolated points. The models in each class of the TOSCA dataset are almost identical in terms of scale, the number of vertices, quality of triangulation, and structure, which all represent the same object with different poses. 

The McGill dataset with articulating parts \cite{siddiqi2008retrieving} is used to evaluate the ability of the descriptors to describe models with poor intra-class quality. The McGill dataset contains 3D models of 30 ants, 30 crabs, 25 glasses, 20 hands, 29 humans, 25 octopuses, 20 pliers, 25 snakes, 31 spiders, and 20 Teddy bears. The classification of the McGill dataset models is more challenging due to scale and shape variations.

\subsection{Retrieval Results}
\label{subsec:retrievalresuls}
In this section, we evaluate the general performance of our proposed shape descriptor and compare it with several state-of-the-art spectral-based global shape descriptors including Shape-DNA~\cite{reuter2006laplace}, cShape-DNA~\cite{gao2014compact}, and GPS~\cite{rustamov2007laplace} algorithms. We chose these methods because they are widely used by researchers (e.g.,~\cite{Mirloo2017,masoumi2017global,boscaini2015learning}) to develop new descriptors or applications, or to evaluate the performance of their proposed descriptors.  Moreover, cShape-DNA represents the normalized version of the original Shape-DNA. Even though there are multiple ways to convert a local point descriptor to a global shape fingerprint, in this article we focus only on algorithms that have been originally introduced as global fingerprints. To this end, we take advantage of the source code made available on Dr. Kokkinos’s homepage\footnote{\url{http://vision.mas.ecp.fr/Personnel/iasonas/descriptors.html}} \cite{bronstein2010scale}, as well as the shape descriptor package provided by Li et al.~\cite{li2014spatially} available on a GitHub repository\footnote{\url{https://github.com/ChunyuanLI/spectral_descriptors}} to generate the Shape-DNA and GPS descriptors, respectively. We also compare the performance of shape retrieval using the code provided for evaluation by SHREC'11~\cite{lian2011shape}. 

The shape descriptors are compared using the TOSCA dataset to discriminate between different classes of 3D objects. In this experiment, we use the first 33 non-zero eigenvalues $(d=33)$. Then, to visualize the locations of objects in the shape space, we project them onto a 2D plane using Principle Component Analysis (PCA). Figure~\ref{fig:2D PCA TOSCA} displays the effectiveness of our method compared with the fingerprints of interest.

\newcommand{\mysize}{0.21}
\begin{figure*}[htb!]
    \centering
    \subfloat[Shape-DNA\label{subfig:tosca-shapedna-2dmds}]{%
    \includegraphics[width=\mysize\textwidth]{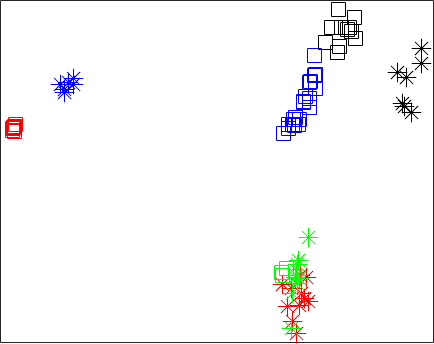}
    } 
    \subfloat[cShape-DNA\label{subfig:tosca-cshapedna-2dmds}]{%
    \includegraphics[width=\mysize\textwidth]{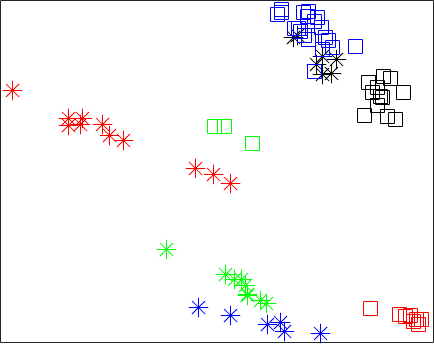}
    } 
    \subfloat[GPS\label{subfig:tosca-gps-2dmds}]{%
    \includegraphics[width=\mysize\textwidth]{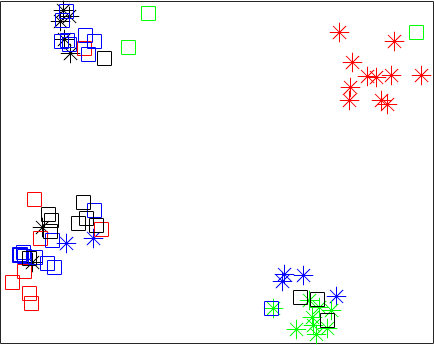}
    }
    \subfloat[LESI\label{subfig:tosca-lesi-2dmds}]{%
    \includegraphics[width=\mysize\textwidth]{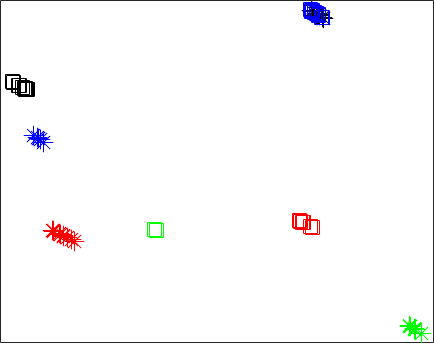}
    } 
    \subfloat
    {\includegraphics[width=0.1\textwidth]{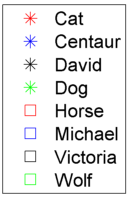}
    }
    \caption{2D PCA projection of shape descriptors computed from \protect\subref{subfig:tosca-shapedna-2dmds} original Shape-DNA, \protect\subref{subfig:tosca-cshapedna-2dmds} cShape-DNA, \protect\subref{subfig:tosca-gps-2dmds} GPS, and \protect\subref{subfig:tosca-lesi-2dmds} LESI algorithms on TOSCA dataset.}
	\label{fig:2D PCA TOSCA}
\end{figure*}

Figure~\ref{fig:2D PCA TOSCA} reveals that LESI can differentiate models of various classes significantly better than the other methods for a refined and normalized dataset. Even though all human models (David, Michael, and Victoria) are very similar, it can distinguish the women from the men's group. However, it fails to discriminate models of Michael from David. Despite the large isometric deformations in each class, the proposed LESI method clusters all models of the same class together very tightly. 

\begin{figure*}[hbt!]
    \centering
    \subfloat[Shape-DNA\label{subfig:mcgill-shapedna-2dmds}]{%
    \includegraphics[width=\mysize\textwidth]{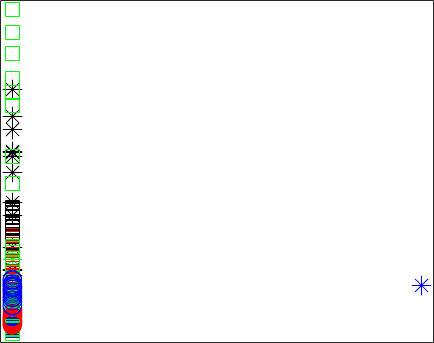}
    }
    \subfloat[cShape-DNA\label{subfig:mcgill-cshapedna-2dmds}]{%
    \includegraphics[width=\mysize\textwidth]{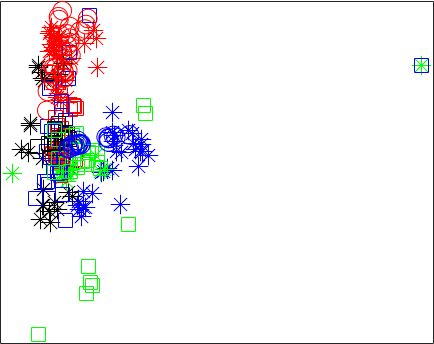}
    }
    \subfloat[GPS\label{subfig:mcgill-gps-2dmds}]{%
    \includegraphics[width=\mysize\textwidth]{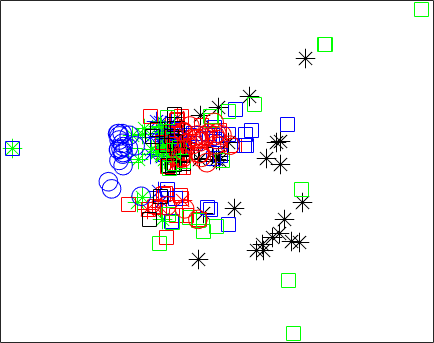}
    }
    \subfloat[LESI\label{subfig:mcgill-lesi-2dmds}]{%
    \includegraphics[width=\mysize\textwidth]{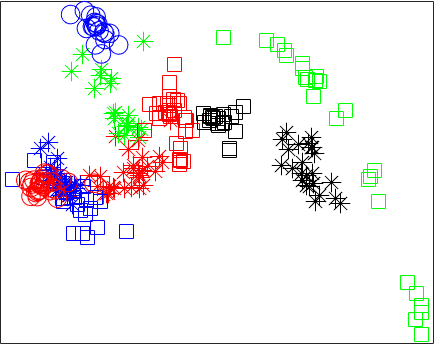}
    }
    \subfloat
    {\includegraphics[width=0.1\textwidth]{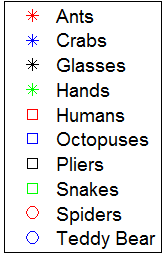}
    }
    \caption{2D PCA projection of shape descriptors computed from \protect\subref{subfig:mcgill-shapedna-2dmds} original Shape-DNA, \protect\subref{subfig:mcgill-cshapedna-2dmds} cShape-DNA, \protect\subref{subfig:mcgill-gps-2dmds} GPS, and \protect\subref{subfig:mcgill-lesi-2dmds} LESI algorithms on McGill dataset.}
	\label{fig:2D PCA McGill}
\end{figure*}

To demonstrate the power of our method in classifying objects with low intra-class similarity compared with other shape descriptors, the same experiment is carried out on the McGill dataset. Models of the same class with articulating parts are in different scales, shape, and structure. The 2D PCA projections of 33-dimension descriptors from all four algorithms are shown in Figure~\ref{fig:2D PCA McGill}.

As illustrated in Figure~\ref{fig:2D PCA McGill}, the original Shape-DNA is highly sensitive to scales. Multiple methods are presented in~\cite{reuter2006laplace} to make the descriptor normalized to scale. cShape-DNA represents a normalized version of it by multiplying the descriptor with the surface area. Although cShape-DNA can separate models from each other, classes are not separated efficiently. LESI outperforms the other algorithms by providing distinct descriptors, which can separate classes. Shape descriptors offered by LESI prove superior to the other algorithms in the shape retrieval and classification tasks, as described below and in the next section respectively.

\begin{table*}[pt!]
\centering
\caption{Shape retrieval performance using TOSCA and McGill datasets}
\label{tbl:retrieval}
\begin{tabular}{l|l|ccccc}
\toprule
Dataset                 & Method    & NN     & FT     & ST     & E      & DCG    \\
\midrule
\multirow{4}{*}{TOSCA}  & ShapeDNA  & \textbf{1.0000} & 0.8091 & 0.9391 & 0.4486 & \textbf{0.9584} \\
                        & cShapeDNA & 0.9474 & 0.7748 & 0.8984 & 0.4748 & 0.9241 \\
                        & GPS       & 0.4868 & 0.4244 & 0.6320 & 0.3614 & 0.6787 \\
                        & LESI      & 0.8684 & \textbf{0.8456} & \textbf{0.9430} & \textbf{0.4860} & 0.9244 \\
\midrule
\multirow{4}{*}{McGill} & ShapeDNA  & 0.7922 & 0.3452 & 0.4977 & 0.3411 & 0.7192 \\
                        & cShapeDNA & 0.7882 & 0.3943 & 0.5483 & 0.3852 & 0.7470 \\
                        & GPS       & 0.3843 & 0.2508 & 0.4066 & 0.2588 & 0.6020 \\
                        & LESI      & \textbf{0.9647} & \textbf{0.7046} & \textbf{0.8739} & \textbf{0.6644} & \textbf{0.9251} \\
\bottomrule
\end{tabular}
\end{table*}

To examine the superiority of LESI quantitatively, we computed multiple standard retrieval measures including Nearest Neighbor (NN), First Tier (FT), Second Tier (ST), e-Measure (E), and Discounted Cumulative Gain (DCG). These measures represent state-of-the-art  quality metrics used when evaluating matching results for shape-based search engines~\cite{shilane2004princeton}. Table~\eqref{tbl:retrieval} reports the results of shape retrieval. Boldface numbers indicate the highest value for each measure per each dataset. From Table~\eqref{tbl:retrieval}, it is clear that the LESI descriptor outperforms all other methods concerning all measures in retrieving models from the McGill dataset. When retrieving models of the TOSCA dataset, LESI outperforms all methods concerning FT, ST, and E measures. Shape-DNA outperforms LESI by a higher value for NN and DCG measures, due to the poor discrimination between David and Michael performed by the LESI descriptor. However, it does not diminish the validity of our claim that LESI performs well for meshes with non-uniform sampling or peculiarities.

\subsection{Multi-class Classification Results}
\label{subsec:classification}
In this section, we corroborate the findings of Section~\ref{subsec:retrievalresuls} by training a linear multi-class SVM classifier to assess the accuracy of LESI compared to other shape descriptors. For this experiment, we utilized the McGill dataset. In addition to the shape descriptors evaluated in Section~\ref{subsec:retrievalresuls}, we computed another normalized version of Shape-DNA by dividing the feature vector by its first element (similar to what LESI offers) as suggested in~\cite{reuter2006laplace}. This way we can compare the effect of the exponential weighting scheme without the influence of the normalization method or compactness (offered by cShape-DNA). Using 10-fold cross-validation and repeating the experiment 3 times, we report the average accuracy for each method in Table~\eqref{tbl:accuracy}. 



The new LESI approach significantly outperforms all other methods when using a two-tailed paired t-test $(p<0.05)$. The t-test was performed on one set of 10 folds in order to avoid violating the independence assumption of the t-test. There is a significant improvement in accuracy when comparing the Shape-DNA (Normalized) to other variants of the Shape-DNA, which is due in part to the normalization method. However, the average accuracy of the LESI descriptor is noticeably higher (95\%) when compared to 90\% of the Shape-DNA (Normalized).

\begin{table}[hbt!]
\centering
\caption{Classification accuracy using McGill dataset}
\label{tbl:accuracy}
\begin{tabular}{lc}
\toprule
\textbf{Method}        & \textbf{Average accuracy} \\
\midrule
Shape-DNA              & 21.02\%    \\
Shape-DNA (Normalized) & 90.60\%    \\
cShape-DNA             & 71.37\%    \\
GPS					   & 50.11\%    \\
\textbf{LESI}          & \textbf{95.69\%}   \\
\bottomrule
\end{tabular}
\end{table}

\begin{figure}[hbt!]
\centering
\includegraphics[width=0.5\textwidth]{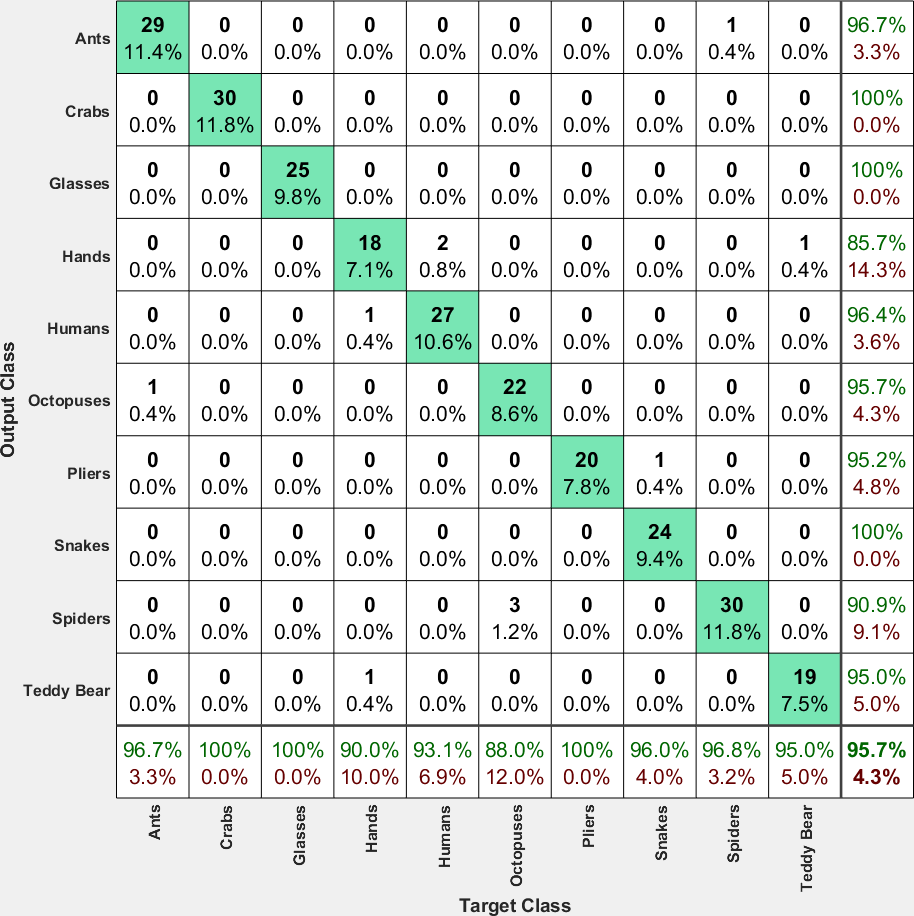}
\caption{Confusion matrix obtained from linear multi-class SVM for McGill dataset using LESI descriptors.}
\label{fig:confmatrix}
\end{figure}

Finally, Figure~\ref{fig:confmatrix} shows the confusion matrix obtained from the linear multi-class SVM using LESI descriptor. The number of correct classifications made for each class (indicated by the green diagonal), confirms that our method captures the discriminative features of the shapes.

\subsection{Robustness}
\label{subsec:robustness}
In this section, we address the robustness of the LESI shape descriptor to shape variations, including noise, scale, and down-sampling by performing another set of experiments. First, we generate the disturbed version of every model in the TOSCA dataset. Then, we test the capability of every method mentioned above in discriminating between different classes. For this purpose, besides plotting the 2D PCA projection of shape descriptors, we also compute and plot the pairwise Euclidean distance matrix, in every case. The distance matrix represents the dissimilarity between each pair of models in the set. It is often used to compute other evaluating metrics such as nearest-neighbor, and first and second tier, to name a few. The dissimilarity of descriptors increases from blue to red, and the more separate classes differ in color, the better they are discriminating from each other.

\begin{figure*}[pbt!]
    \centering
    \subfloat
    {\includegraphics[width=\mysize\textwidth]{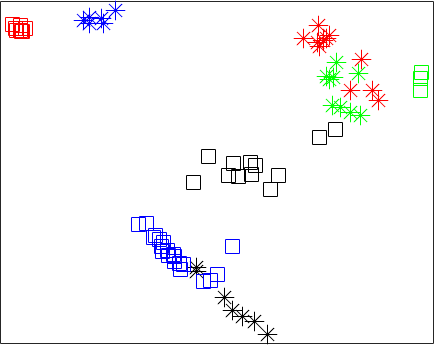}
    }
    \subfloat
    {\includegraphics[width=\mysize\textwidth]{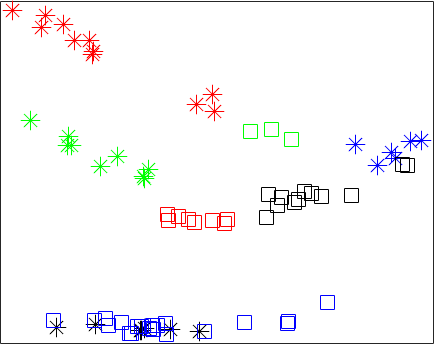}
    }
    \subfloat
    {\includegraphics[width=\mysize\textwidth]{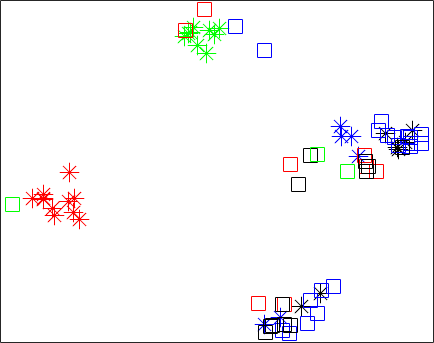}
    }
    \subfloat
    {\includegraphics[width=\mysize\textwidth]{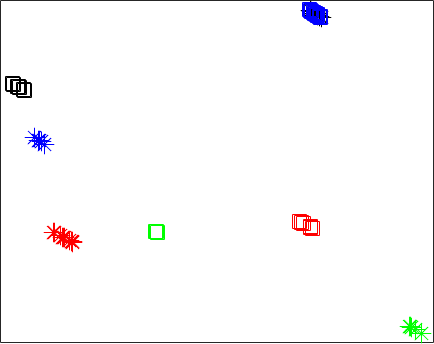}
    }
    \subfloat
    {\includegraphics[width=0.1\textwidth]{Images/legend_TOSCA.png}}
    \\
    \subfloat
    {\includegraphics[width=\mysize\textwidth]{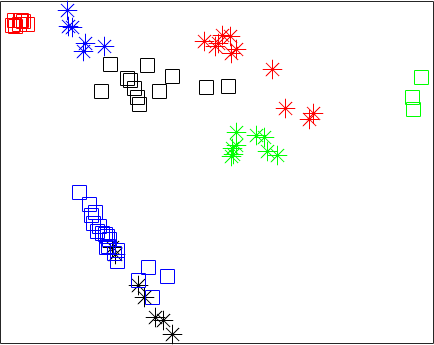}
    }
    \subfloat
    {\includegraphics[width=\mysize\textwidth]{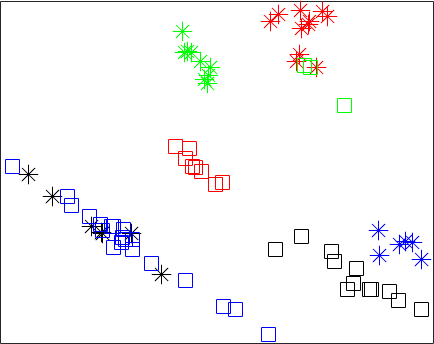}
    }
    \subfloat
    {\includegraphics[width=\mysize\textwidth]{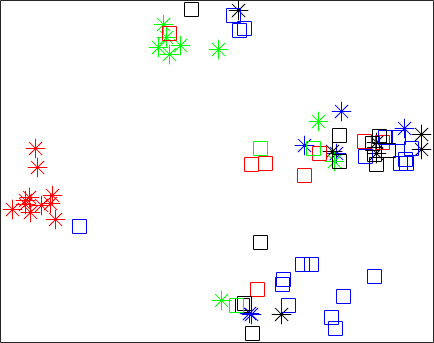}
    }
    \subfloat
    {\includegraphics[width=\mysize\textwidth]{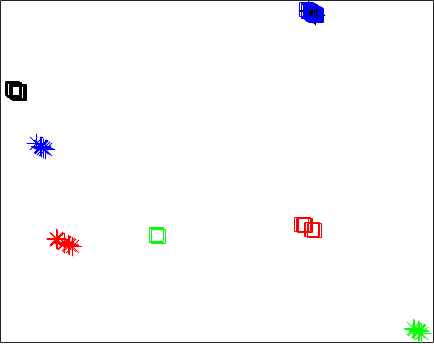}
    }
    \subfloat
    {\includegraphics[width=0.1\textwidth]{Images/legend_TOSCA.png}}
    \\
    \subfloat
    {\includegraphics[width=\mysize\textwidth]{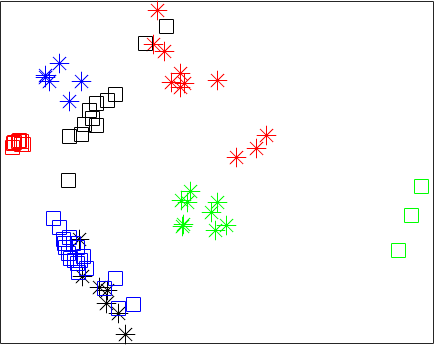}
    }
    \subfloat
    {\includegraphics[width=\mysize\textwidth]{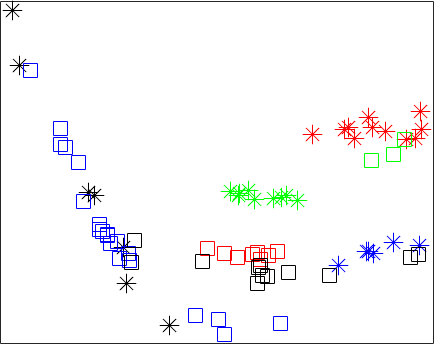}
    }
    \subfloat
    {\includegraphics[width=\mysize\textwidth]{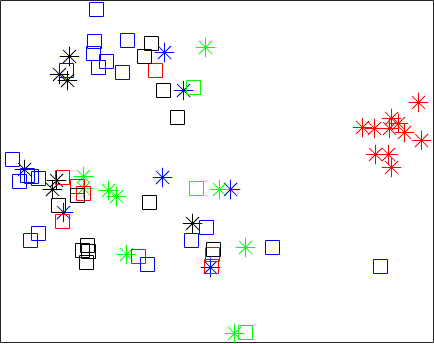}
    }
    \subfloat
    {\includegraphics[width=\mysize\textwidth]{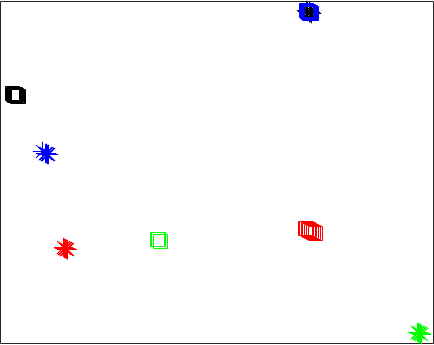}
    }
    \subfloat
    {\includegraphics[width=0.1\textwidth]{Images/legend_TOSCA.png}}
    \caption{2D PCA projection of shape descriptors computed by (from left to right) Shape-DNA, cShape-DNA, GPS, and LESI algorithms from perturbed TOSCA dataset with (from top to bottom) 0.5\%, 1\%, 2\% noise level, respectively.}
	\label{fig:Robustness to noise 2D PCA}
\end{figure*}

\textbf{Resistance to noise.} Multiple noisy versions of the TOSCA dataset are generated following the idea articulated in~\cite{liu2011construction}. To this end, the surface meshes of all models are disturbed by changing the position of each point along its normal vector that is chosen randomly from an interval $(-L, L)$ with the 0 mean, where $L$ determines the noise level and is a fraction of the diagonal length of the model bounding box. In this experiment, three noise levels $L=0.5\%$, $L=1\%$, and $L=2\%$ are tested, where the latter one represents a greater level of noise. Two-dimensional PCA projections of all descriptors with the presence of different levels of noise are plotted in Figure~\protect\ref{fig:Robustness to noise 2D PCA}. Combining these with the results shown in Figure~\ref{fig:2D PCA TOSCA}, where no noise is present, demonstrates that the LESI algorithm is highly noise-resistant while the performance of the Shape-DNA and cShape-DNA decreases as the level of noise increases. Moreover, GPS fails in separating different classes of models with the presence of noise. Figure~\ref{fig:Robustness to noise DistMatrix} reflects the effect of noise on the discriminative power of the descriptors. The LESI algorithm shows consistent results as the level of noise increases from 0\% (top row) to 2\% (bottom row).

\begin{figure*}[pbt!]
    \centering
    \subfloat[Shape-DNA\label{subfig:tosca-rndscale-shapedna-2dmds}]
    {\includegraphics[width=\mysize\textwidth]{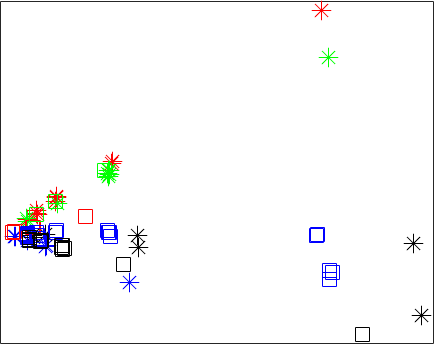}
    }
    \subfloat[cShape-DNA\label{subfig:tosca-rndscale-cshapedna-2dmds}]
    {\includegraphics[width=\mysize\textwidth]{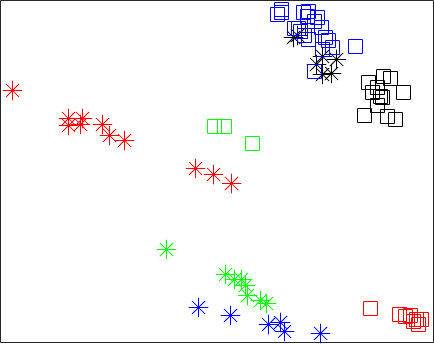}
    }
    \subfloat[GPS\label{subfig:tosca-rndscale-gps-2dmds}]
    {\includegraphics[width=\mysize\textwidth]{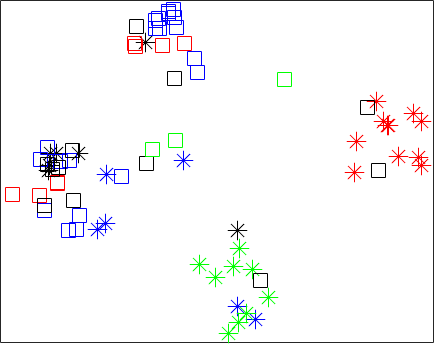}
    }
    \subfloat[LESI\label{subfig:tosca-rndscale-lesi-2dmds}]
    {\includegraphics[width=\mysize\textwidth]{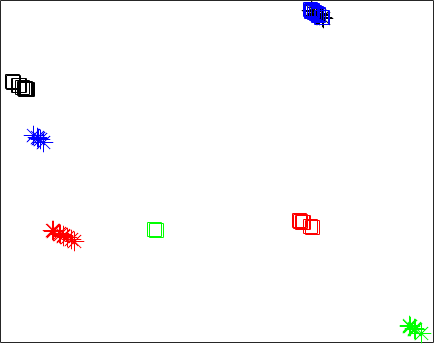}
    }
    \subfloat
    {\includegraphics[width=0.1\textwidth]{Images/legend_TOSCA.png}
    }
    \caption{2D PCA projection of shape descriptors computed by \protect\subref{subfig:tosca-rndscale-shapedna-2dmds} original Shape-DNA, \protect\subref{subfig:tosca-rndscale-cshapedna-2dmds} cShape-DNA, \protect\subref{subfig:tosca-rndscale-gps-2dmds} GPS, and \protect\subref{subfig:tosca-rndscale-lesi-2dmds} LESI algorithms over scaled version of the TOSCA dataset by a randomly chosen factor of 0.5, 0.875, 1.25, 1.625, or 2.}
	\label{fig:Robustness to scale 2D PCA}
\end{figure*}

\textbf{Scale invariance.} In order to validate the insensitivity of the LESI descriptor to scale variations and compare the robustness of the proposed method with other descriptors, each model of the TOSCA dataset is scaled by a factor of 0.5, 0.875, 1.25, 1.625, or 2 randomly. Figure~\ref{fig:Robustness to scale 2D PCA} shows that the LESI algorithm surpasses other methods in discerning different classes. Comparing the result of this experiment with the results shown in Figure~\ref{fig:2D PCA TOSCA} demonstrates the consistency of the LESI algorithm with the presence of scale variation. The distance matrices in Figure~\ref{fig:Robustness to scale DistMatrix} show that the original Shape-DNA algorithm is very susceptible to scale variations. Even though the cShape-DNA has significantly improved scale sensitivity of the original Shape-DNA, it does not provide as accurate results as the LESI algorithm does.

\begin{figure*}[pbt!]
    \centering
    \subfloat[Shape-DNA\label{subfig:tosca-dwnsmpl20-shapedna-2dmds}]
    {\includegraphics[width=\mysize\textwidth]{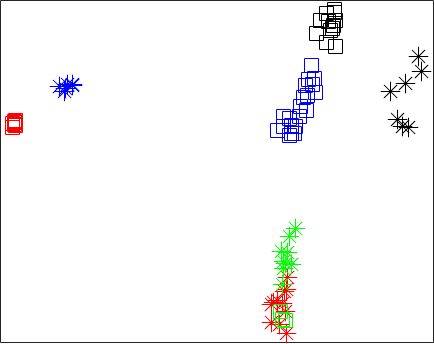}
    }
    \subfloat[cShape-DNA\label{subfig:tosca-dwnsmpl20-cshapedna-2dmds}]
    {\includegraphics[width=\mysize\textwidth]{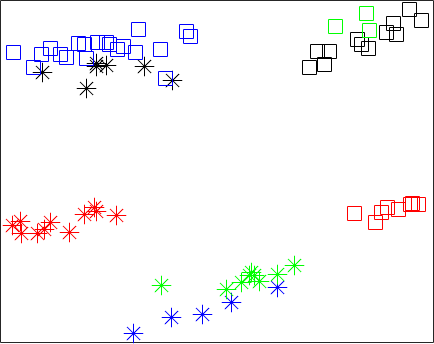}
    }
    \subfloat[GPS\label{subfig:tosca-dwnsmpl20-gps-2dmds}]
    {\includegraphics[width=\mysize\textwidth]{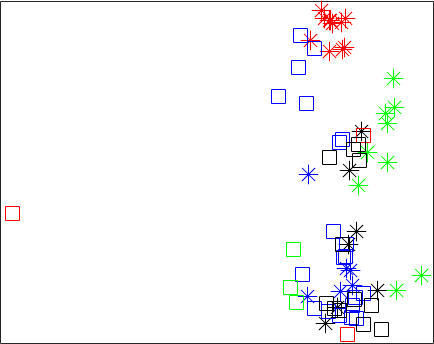}
    }
    \subfloat[LESI\label{subfig:tosca-dwnsmpl20-lesi-2dmds}]
    {\includegraphics[width=\mysize\textwidth]{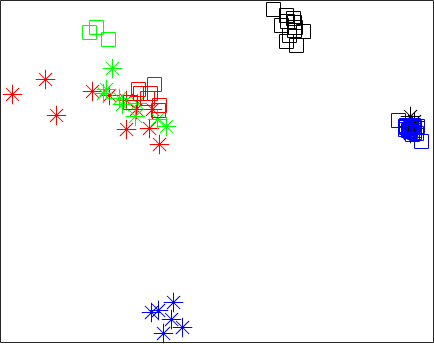}
    }
    \subfloat
    {\includegraphics[width=0.1\textwidth]{Images/legend_TOSCA.png}
    }
    \caption{2D PCA projection of shape descriptors computed by \protect\subref{subfig:tosca-dwnsmpl20-shapedna-2dmds} original Shape-DNA, \protect\subref{subfig:tosca-dwnsmpl20-cshapedna-2dmds} cShape-DNA, \protect\subref{subfig:tosca-dwnsmpl20-gps-2dmds} GPS, and \protect\subref{subfig:tosca-dwnsmpl20-lesi-2dmds} LESI algorithms from down sampled TOSCA dataset by rate of 20\%.}
	\label{fig:Robustness to sampling rate 2D PCA}
\end{figure*}

\textbf{Resistance to the sampling rate.}
To investigate the effect of sampling rates on the discriminative power of the shape descriptors, Bronstein et al.~\cite{bronstein2011shape} propose to reduce the number of vertices to 20\% of its original size. Accordingly, the down-sampled version of the TOSCA dataset is generated, and shape descriptors associated with them are computed. The 2D PCA projections and distance matrices of descriptors are illustrated in Figures~\ref{fig:Robustness to sampling rate 2D PCA} and~\ref{fig:Robustness to sampling rate DistMatrix}, respectively. Although the original Shape-DNA shows a more accurate result than cShape-DNA, the separation of cat, dog, and wolf models is challenging. Although the performance of the LESI method is slightly affected, it still outperforms cShape-DNA and GPS methods.

\renewcommand{\mysize}{0.18}
\begin{figure*}[hbt!]
    \centering
	\resizebox{\textwidth}{!}{
	\begin{tabular}{c c c c c}
	\toprule
	\begin{tabular}{@{}c@{}}\textbf{Noise} \\ \textbf{Level}\end{tabular} & \textbf{Shape-DNA} & \textbf{cShape-DNA} & \textbf{GPS} & \textbf{LESI}\\
	\midrule
    \textbf{0\%} &
    \begin{tabular}[c]{@{}l@{}}\includegraphics[width=\mysize\textwidth]{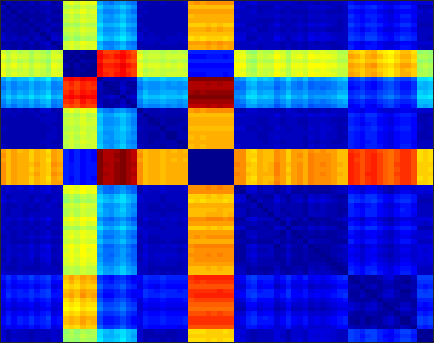}\end{tabular} &
    \begin{tabular}[c]{@{}l@{}}\includegraphics[width=\mysize\textwidth]{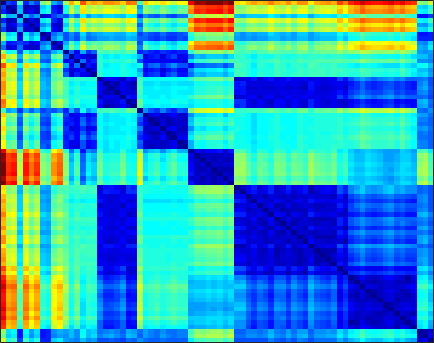}\end{tabular} &
    \begin{tabular}[c]{@{}l@{}}\includegraphics[width=\mysize\textwidth]{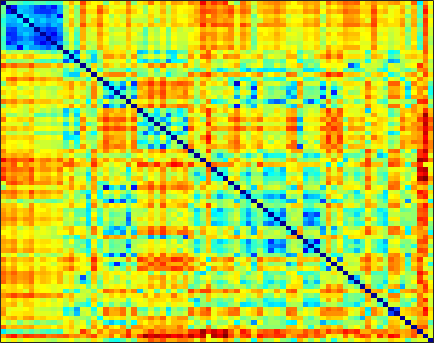}\end{tabular} &
    \begin{tabular}[c]{@{}l@{}}\includegraphics[width=\mysize\textwidth]{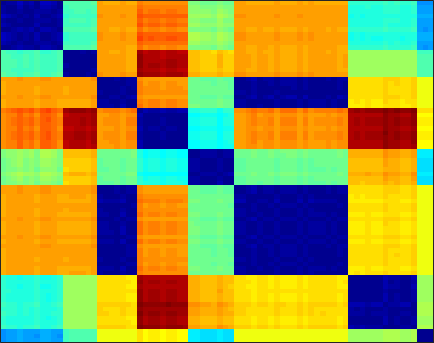}\end{tabular} \\
    \textbf{0.5}\% &
    \begin{tabular}[c]{@{}l@{}}\includegraphics[width=\mysize\textwidth]{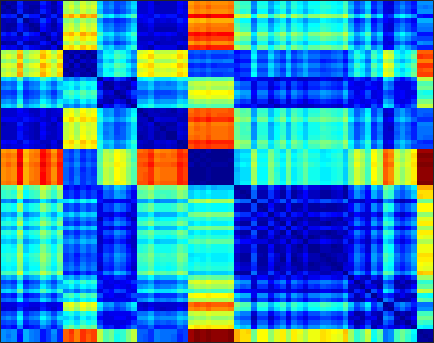}\end{tabular} &
    \begin{tabular}[c]{@{}l@{}}\includegraphics[width=\mysize\textwidth]{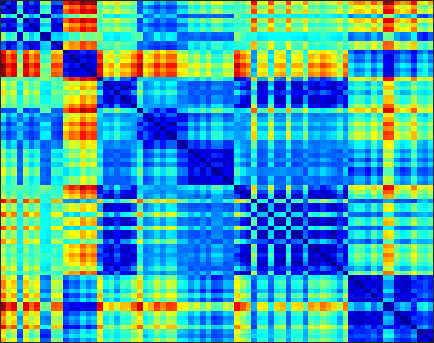}\end{tabular} &
    \begin{tabular}[c]{@{}l@{}}\includegraphics[width=\mysize\textwidth]{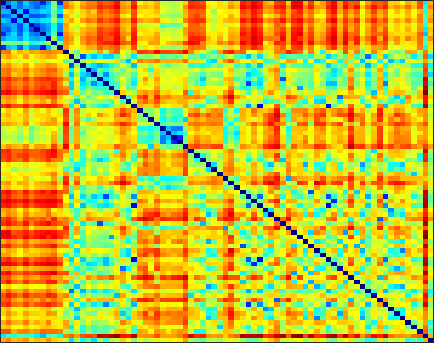}\end{tabular} &
    \begin{tabular}[c]{@{}l@{}}\includegraphics[width=\mysize\textwidth]{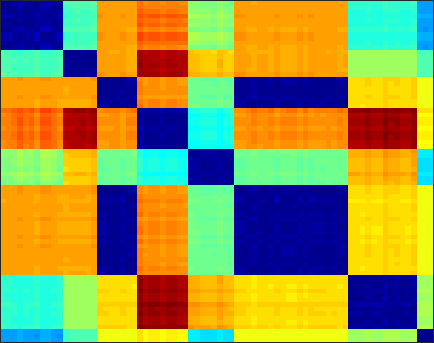}\end{tabular} \\
    \textbf{1\%} &
    \begin{tabular}[c]{@{}l@{}}\includegraphics[width=\mysize\textwidth]{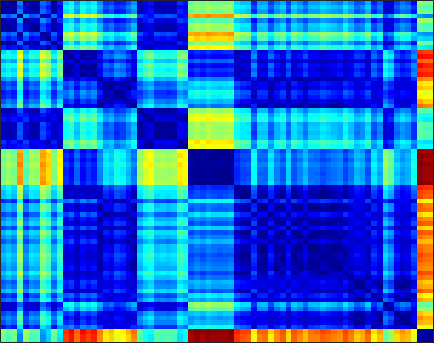}\end{tabular} &
    \begin{tabular}[c]{@{}l@{}}\includegraphics[width=\mysize\textwidth]{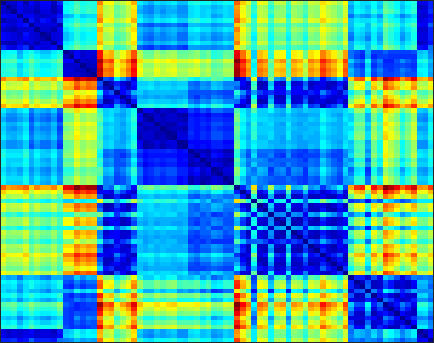}\end{tabular} &
    \begin{tabular}[c]{@{}l@{}}\includegraphics[width=\mysize\textwidth]{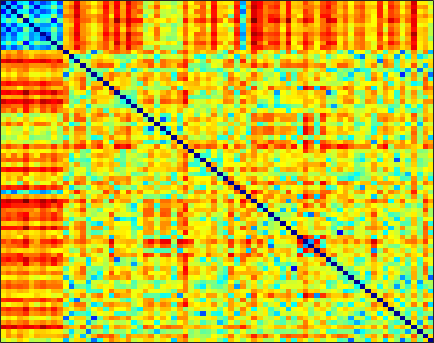}\end{tabular} &
    \begin{tabular}[c]{@{}l@{}}\includegraphics[width=\mysize\textwidth]{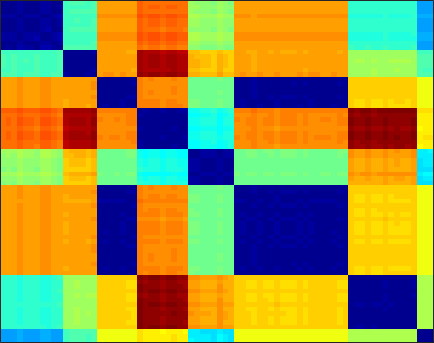}\end{tabular} \\
    \textbf{2\%} &
    \begin{tabular}[c]{@{}l@{}}\includegraphics[width=\mysize\textwidth]{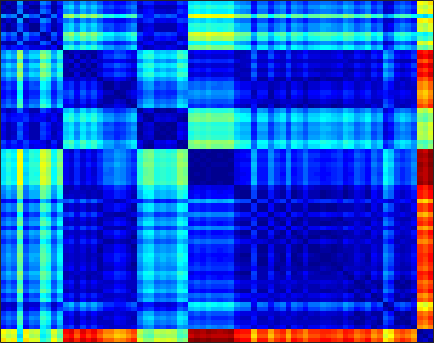}\end{tabular} &
    \begin{tabular}[c]{@{}l@{}}\includegraphics[width=\mysize\textwidth]{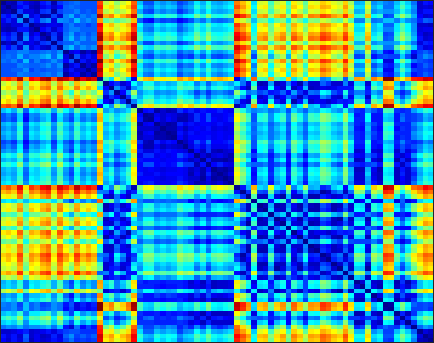}\end{tabular} &
    \begin{tabular}[c]{@{}l@{}}\includegraphics[width=\mysize\textwidth]{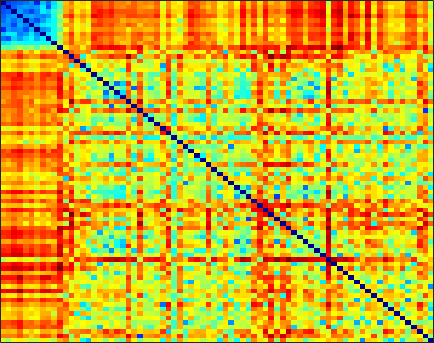}\end{tabular} &
    \begin{tabular}[c]{@{}l@{}}\includegraphics[width=\mysize\textwidth]{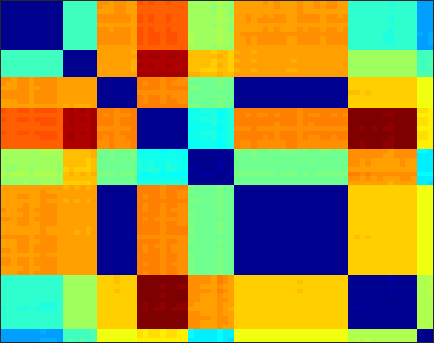}\end{tabular} \\
    \bottomrule
	\end{tabular}}
    \caption{The Euclidean pairwise distance matrix of shape descriptors computed by (from left to right) Shape-DNA, cShape-DNA, GPS, and LESI algorithms from perturbed TOSCA dataset by (from top to bottom) 0\%, 0.5\%, 1\%, 2\% noise levels.}
	\label{fig:Robustness to noise DistMatrix}
\end{figure*}

\begin{figure*}[pbt!]
	\centering
   	\resizebox{\textwidth}{!}{
	\begin{tabular}{c c c c c}
	\toprule
	 & \textbf{Shape-DNA} & \textbf{cShape-DNA} & \textbf{GPS} & \textbf{LESI}\\
	\midrule
    \begin{tabular}{@{}c@{}}\textbf{Random} \\ \textbf{Scale}\end{tabular} &
    \begin{tabular}[c]{@{}l@{}}\includegraphics[width=\mysize\textwidth]{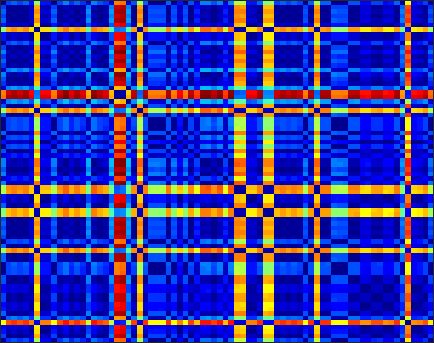}
    \end{tabular} &
    \begin{tabular}[c]{@{}l@{}}\includegraphics[width=\mysize\textwidth]{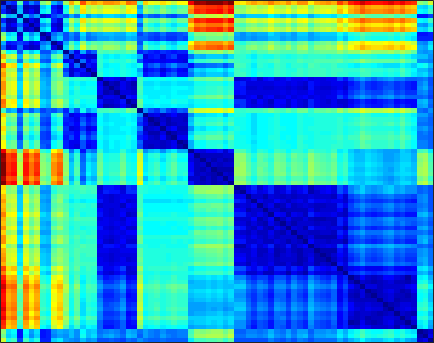}
    \end{tabular} &
    \begin{tabular}[c]{@{}l@{}}\includegraphics[width=\mysize\textwidth]{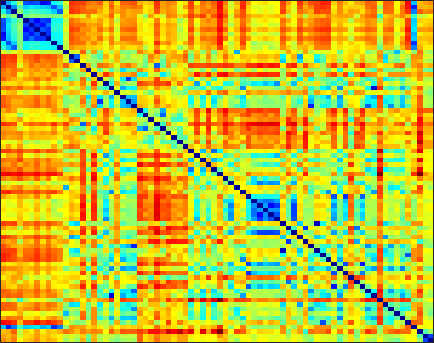}
    \end{tabular} &
    \begin{tabular}[c]{@{}l@{}}\includegraphics[width=\mysize\textwidth]{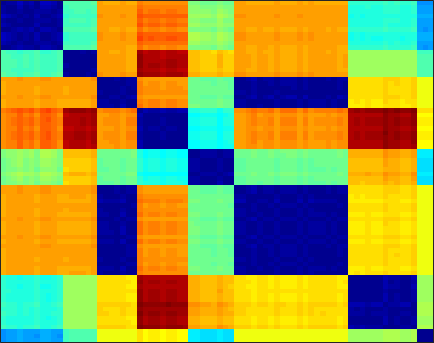}
    \end{tabular} \\
    \bottomrule
	\end{tabular}}
    \caption{The Euclidean pairwise distance matrix of shape descriptors computed by (from left to right) Shape-DNA, cShape-DNA, GPS, and LESI algorithms over scaled version of the TOSCA dataset by a randomly chosen factor of 0.5, 0.875, 1.25, 1.625, or 2.}
	\label{fig:Robustness to scale DistMatrix}
\end{figure*}

\begin{figure*}[pbt!]
    \centering
    \resizebox{\textwidth}{!}{
	\begin{tabular}{c c c c c}
	\toprule
	 & \textbf{Shape-DNA} & \textbf{cShape-DNA} & \textbf{GPS} & \textbf{LESI}\\
	\midrule
    \begin{tabular}{@{}c@{}}\textbf{Down} \\ \textbf{Sampled}\end{tabular} &
    \begin{tabular}[c]{@{}l@{}}\includegraphics[width=\mysize\textwidth]{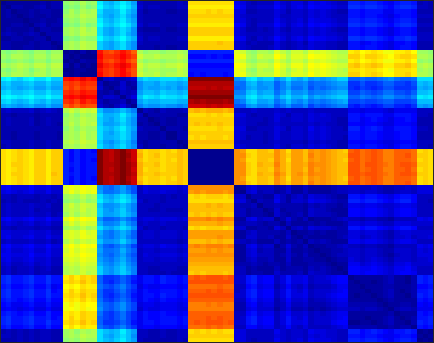}
    \end{tabular} &
    \begin{tabular}[c]{@{}l@{}}\includegraphics[width=\mysize\textwidth]{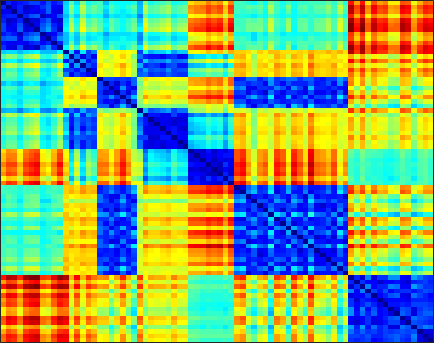}
    \end{tabular} &
    \begin{tabular}[c]{@{}l@{}}\includegraphics[width=\mysize\textwidth]{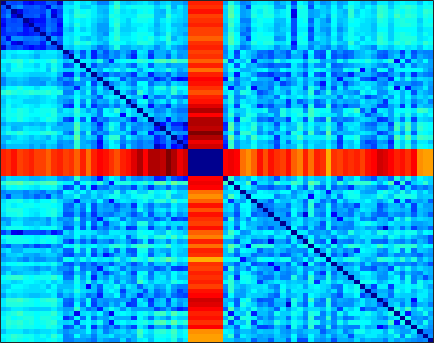}
    \end{tabular} &
    \begin{tabular}[c]{@{}l@{}}\includegraphics[width=\mysize\textwidth]{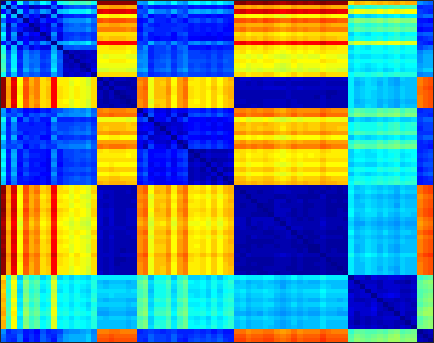}
    \end{tabular} \\
    \bottomrule
	\end{tabular}}
    \caption{The Euclidean pairwise distance matrix of shape descriptors computed by (from left to right) Shape-DNA, cShape-DNA, GPS, and LESI algorithms from from down sampled TOSCA dataset by rate of 20\%.}
	\label{fig:Robustness to sampling rate DistMatrix}
\end{figure*}

\section{Discussion}
\label{sec:discuss}
In this article, motivated by the unique properties of Laplacian Eigenmap (i.e., locality preservation, structural equivalence, and dimensionality reduction) and inspired by the existing spectral-based shape descriptors, we investigated the application of manifold learning in deriving a shape fingerprint in order to address the limitations tied to popular cotangent-based shape descriptors. We proposed a global descriptor (LESI) with an easy-to-compute and efficient normalization technique that facilitates applications such as shape classification and retrieval. Our method applies fewer restrictions on the class of meshes as well as improving the quality of tessellations. Analogous to other spectral descriptors, LESI uses the spectrum of the LB operator, which is independent of the shape location, is informative (contains a considerable amount of geometrical and topological information), and above all isometric invariant. We compared the discriminating power of LESI with three prominent descriptors from the literature, namely Shape-DNA, cShape-DNA, and GPS, and found it to be superior. 

In the first set of experiments illustrated in Figures~\ref{fig:2D PCA TOSCA} and \ref{fig:2D PCA McGill}, our method substantially outperforms the others. The superiority of LESI is more significant when the McGill dataset is used (Table~\ref{tbl:accuracy} and Figure\ref{fig:confmatrix}). This dataset includes wide variations in mesh structure and scales, causing the failure of the other methods to generate acceptable results. However, LESI, due to utilizing a different method of discretization to form the LB operator, focuses on the vicinity rather than the quality of the triangulation. Therefore, our technique, unlike other methods, is not affected by the low quality of polygon meshes.

The second set of experiments evaluates the reliability of our method in the presence of noise, scale variations, as well as different sampling rates. LESI shows impressive robustness against the first two sets of perturbation. Despite the negative impact of down sampling in LESI descriptor, it continues to show better performance when compared to cShape-DNA and GPS. It should be noted that the result could also be improved by increasing the size of the output vector.

In addition to the discriminating power of the descriptor, degenerate and non-uniform meshes may also cause failure of an algorithm to converge. The cotangent weight-based algorithms were not able to compute the descriptors for 2 shapes from the McGill dataset. GPS also failed to compute descriptors for 6 models of the down sampled TOSCA dataset. However, our technique converges at all times despite the quality of the polygon mesh structure. 

Moreover, LESI, unlike cotangent weight-based techniques, is not confined to the triangulated meshes as it disregards the mesh geometry~\cite{zhang2009surface}. LESI inherits this property from the capability of manifold learning techniques in coping with high dimensional data. The discretization of the LB operator using cotangent weights on the quadrilateral meshes is not as straightforward as on triangular meshes. To compute the LB operator on a quadrilateral mesh, all rectangles need to be divided into triangles. It could be done easily, however, as for each quad there are two possible triangulations, the result is not unique. 

In the original Laplacian Eigenmaps, the high dimension data requires a considerable amount of processing as the list of all connections need to be computed for the dataset. In fact, for each point in the high dimension space, a given number of nearest neighbors need to be extracted which could be challenging and unmanageable. While applying this technique to the 3D meshes, we skip this step as the neighbors are already defined and given in the mesh structure. 

This work benefits from the Laplacian Eigenmap technique in a space in which the vicinities are given. LESI takes advantage of simple Laplacian computation, to form the LB operator, which provides concise and informative shape descriptors. Experimental results prove that LESI is more effective compared with the other powerful descriptors.

One limitation of LESI is its inability to separate models of different men (David and Michael). However, it was able to differentiate between the wolf and dog, as well as between women and men. 

Although we investigated only the application of Laplacian Eigenmap in introducing a shape descriptor, there are some other spectral-based manifold learning methods, such as Isomap, LLE, and Diffusion map, which have not been examined. This can be considered future work.

\section{Conclusion}
\label{sec:conclusion}

This work presents LESI, a novel scale-invariant global shape descriptor based on Laplacian Eigenmap that is significantly better when compared to other shape descriptors. We conclude that manifold learning methods can be used to develop new spectral-based shape descriptors to learn the structure of manifolds despite the quality of sampled meshes.

\section{Acknowledgment}
We acknowledge financial support for Drs. D'Souza, Yu and Ms. Bashiri from GE Healthcare through the UWM Catalyst Grant program. Our sincere thanks goes to Dr. Ahmad P. Tafti and Dr. C. David Page for their expertise and constructive comments. We also acknowledge financial support from the Center for Predictive Computational Phenotyping, supported by the National Institutes of Health Big Data to Knowledge (BD2K) Initiative under Award Number U54 AI117924 and the grant UL1TR002373 from the Clinical and Translational Science Award (CTSA) program of the National Center for Advancing Translational Sciences, NIH.

\bibliographystyle{unsrtnat}
\bibliography{sample.bib}

\begin{thebibliography}{54}
\providecommand{\natexlab}[1]{#1}
\providecommand{\url}[1]{\texttt{#1}}
\expandafter\ifx\csname urlstyle\endcsname\relax
  \providecommand{\doi}[1]{doi: #1}\else
  \providecommand{\doi}{doi: \begingroup \urlstyle{rm}\Url}\fi

\bibitem[Omrani et~al.(2016)Omrani, Tafti, Fathi, Moghadam, Rohatgi, D'Souza,
  and Yu]{omrani16}
Emad Omrani, Ahmad~P Tafti, Mojtaba~F Fathi, Afsaneh~Dorri Moghadam, Pradeep
  Rohatgi, Roshan~M D'Souza, and Zeyun Yu.
\newblock Tribological study in microscale using 3d sem surface reconstruction.
\newblock \emph{Tribology International}, 103:\penalty0 309--315, 2016.

\bibitem[Ng et~al.(2007)Ng, Pathak, Kuan, Lau, Dong, Sodt, Dang, Avants,
  Yushkevich, Gee, et~al.]{ng2007neuroinformatics}
Lydia Ng, Sayan Pathak, Chihchau Kuan, Chris Lau, Hong-wei Dong, Andrew Sodt,
  Chinh Dang, Brian Avants, Paul Yushkevich, James Gee, et~al.
\newblock Neuroinformatics for genome-wide 3-d gene expression mapping in the
  mouse brain.
\newblock \emph{IEEE/ACM Transactions on Computational Biology and
  Bioinformatics (TCBB)}, 4\penalty0 (3):\penalty0 382--393, 2007.

\bibitem[Gao et~al.(2016)Gao, Rostami, Pang, Fu, and Yu]{gao2016mesh}
Zhanheng Gao, Reihaneh Rostami, Xiaoli Pang, Zhicheng Fu, and Zeyun Yu.
\newblock Mesh generation and flexible shape comparisons for bio-molecules.
\newblock \emph{Molecular Based Mathematical Biology}, 4\penalty0 (1):\penalty0
  1--13, 2016.

\bibitem[Sosa et~al.(2016)Sosa, Rodríguez, Guaje, Victorino, Mejía, Fuentes,
  Ramírez, and Franco]{Sosa20163d7743319}
G.~D. Sosa, S.~Rodríguez, J.~Guaje, J.~Victorino, M.~Mejía, L.~S. Fuentes,
  A.~Ramírez, and H.~Franco.
\newblock 3d surface reconstruction of entomological specimens from uniform
  multi-view image datasets.
\newblock In \emph{2016 XXI Symposium on Signal Processing, Images and
  Artificial Vision (STSIVA)}, pages 1--8, Aug 2016.
\newblock \doi{10.1109/STSIVA.2016.7743319}.

\bibitem[Riehemann et~al.(2011)Riehemann, Palme, Kuehmstedt, Grossmann, Notni,
  and Hintersehr]{riehemann2011microdisplay}
Stefan Riehemann, Martin Palme, Peter Kuehmstedt, Constanze Grossmann, Gunther
  Notni, and Josef Hintersehr.
\newblock Microdisplay-based intraoral 3d scanner for dentistry.
\newblock \emph{Journal of Display Technology}, 7\penalty0 (3):\penalty0
  151--155, 2011.

\bibitem[Wu et~al.(2016)Wu, Bradley, Garrido, Zollh{\"o}fer, Theobalt, Gross,
  and Beeler]{wu2016model}
Chenglei Wu, Derek Bradley, Pablo Garrido, Michael Zollh{\"o}fer, Christian
  Theobalt, Markus Gross, and Thabo Beeler.
\newblock Model-based teeth reconstruction.
\newblock \emph{ACM Transactions on Graphics (TOG)}, 35\penalty0 (6):\penalty0
  220, 2016.

\bibitem[Aflalo et~al.(2011)Aflalo, Bronstein, Bronstein, and
  Kimmel]{aflalo2011deformable}
Yonathan Aflalo, Alexander~M Bronstein, Michael~M Bronstein, and Ron Kimmel.
\newblock Deformable shape retrieval by learning diffusion kernels.
\newblock In \emph{International Conference on Scale Space and Variational
  Methods in Computer Vision}, pages 689--700. Springer, 2011.

\bibitem[Bronstein et~al.(2011)Bronstein, Bronstein, Guibas, and
  Ovsjanikov]{bronstein2011shape}
Alexander~M Bronstein, Michael~M Bronstein, Leonidas~J Guibas, and Maks
  Ovsjanikov.
\newblock Shape google: Geometric words and expressions for invariant shape
  retrieval.
\newblock \emph{ACM Transactions on Graphics (TOG)}, 30\penalty0 (1):\penalty0
  1, 2011.

\bibitem[Xie et~al.(2015)Xie, Fang, Zhu, and Wong]{xie2015deepshape}
Jin Xie, Yi~Fang, Fan Zhu, and Edward Wong.
\newblock Deepshape: Deep learned shape descriptor for 3d shape matching and
  retrieval.
\newblock In \emph{Proceedings of the IEEE Conference on Computer Vision and
  Pattern Recognition}, pages 1275--1283, 2015.

\bibitem[Toldo et~al.(2009)Toldo, Castellani, and Fusiello]{toldo2009visual}
R~Toldo, U~Castellani, and A~Fusiello.
\newblock Visual vocabulary signature for 3d object retrieval and partial
  matching.
\newblock In \emph{Proceedings of the 2nd Eurographics conference on 3D Object
  Retrieval}, pages 21--28. Eurographics Association, 2009.

\bibitem[Bu et~al.(2014)Bu, Liu, Han, Wu, and Ji]{bu2014learning}
Shuhui Bu, Zhenbao Liu, Junwei Han, Jun Wu, and Rongrong Ji.
\newblock Learning high-level feature by deep belief networks for 3-d model
  retrieval and recognition.
\newblock \emph{IEEE Transactions on Multimedia}, 16\penalty0 (8):\penalty0
  2154--2167, 2014.

\bibitem[Boscaini et~al.(2016)Boscaini, Masci, Rodol{\`a}, Bronstein, and
  Cremers]{boscaini2016anisotropic}
Davide Boscaini, Jonathan Masci, Emanuele Rodol{\`a}, Michael~M Bronstein, and
  Daniel Cremers.
\newblock Anisotropic diffusion descriptors.
\newblock \emph{Computer Graphics Forum}, 35\penalty0 (2):\penalty0 431--441,
  2016.

\bibitem[Bronstein(2011)]{bronstein2011spectral}
Alexander~M Bronstein.
\newblock Spectral descriptors for deformable shapes.
\newblock \emph{arXiv preprint arXiv:1110.5015}, 2011.

\bibitem[Raviv et~al.(2013)Raviv, Kimmel, and Bruckstein]{raviv2013graph}
Dan Raviv, Ron Kimmel, and Alfred~M Bruckstein.
\newblock Graph isomorphisms and automorphisms via spectral signatures.
\newblock \emph{IEEE transactions on pattern analysis and machine
  intelligence}, 35\penalty0 (8):\penalty0 1985--1993, 2013.

\bibitem[De~Youngster et~al.(2013)De~Youngster, Paquet, Viktor, and
  Petriu]{de2013isometry}
Dela De~Youngster, Eric Paquet, Herna Viktor, and Emil Petriu.
\newblock An isometry-invariant spectral approach for protein-protein docking.
\newblock In \emph{Bioinformatics and Bioengineering (BIBE), 2013 IEEE 13th
  International Conference on}, pages 1--6. IEEE, 2013.

\bibitem[Ovsjanikov et~al.(2012)Ovsjanikov, Ben-Chen, Solomon, Butscher, and
  Guibas]{ovsjanikov2012functional}
Maks Ovsjanikov, Mirela Ben-Chen, Justin Solomon, Adrian Butscher, and Leonidas
  Guibas.
\newblock Functional maps: a flexible representation of maps between shapes.
\newblock \emph{ACM Transactions on Graphics (TOG)}, 31\penalty0 (4):\penalty0
  30, 2012.

\bibitem[Reuter et~al.(2006)Reuter, Wolter, and Peinecke]{reuter2006laplace}
Martin Reuter, Franz-Erich Wolter, and Niklas Peinecke.
\newblock Laplace--beltrami spectra as 'shape-dna' of surfaces and solids.
\newblock \emph{Computer-Aided Design}, 38\penalty0 (4):\penalty0 342--366,
  2006.

\bibitem[Sun et~al.(2009)Sun, Ovsjanikov, and Guibas]{sun2009concise}
Jian Sun, Maks Ovsjanikov, and Leonidas Guibas.
\newblock A concise and provably informative multi-scale signature based on
  heat diffusion.
\newblock \emph{Computer graphics forum}, 28\penalty0 (5):\penalty0 1383--1392,
  2009.

\bibitem[Aubry et~al.(2011)Aubry, Schlickewei, and Cremers]{aubry2011wave}
Mathieu Aubry, Ulrich Schlickewei, and Daniel Cremers.
\newblock The wave kernel signature: A quantum mechanical approach to shape
  analysis.
\newblock In \emph{Computer Vision Workshops (ICCV Workshops), 2011 IEEE
  International Conference on}, pages 1626--1633. IEEE, 2011.

\bibitem[Rustamov(2007)]{rustamov2007laplace}
Raif~M Rustamov.
\newblock Laplace-beltrami eigenfunctions for deformation invariant shape
  representation.
\newblock In \emph{Proceedings of the fifth Eurographics symposium on Geometry
  processing}, pages 225--233. Eurographics Association, 2007.

\bibitem[Reuter et~al.(2009)Reuter, Biasotti, Giorgi, Patan{\`e}, and
  Spagnuolo]{reuter2009discrete}
Martin Reuter, Silvia Biasotti, Daniela Giorgi, Giuseppe Patan{\`e}, and
  Michela Spagnuolo.
\newblock Discrete laplace--beltrami operators for shape analysis and
  segmentation.
\newblock \emph{Computers \& Graphics}, 33\penalty0 (3):\penalty0 381--390,
  2009.

\bibitem[Belkin et~al.(2008)Belkin, Sun, and Wang]{belkin2008discrete}
Mikhail Belkin, Jian Sun, and Yusu Wang.
\newblock Discrete laplace operator on meshed surfaces.
\newblock In \emph{Proceedings of the twenty-fourth annual symposium on
  Computational geometry}, pages 278--287. ACM, 2008.

\bibitem[Goldberg et~al.(2008)Goldberg, Zakai, Kushnir, and
  Ritov]{goldberg2008manifold}
Yair Goldberg, Alon Zakai, Dan Kushnir, and Ya’acov Ritov.
\newblock Manifold learning: The price of normalization.
\newblock \emph{Journal of Machine Learning Research}, 9\penalty0
  (Aug):\penalty0 1909--1939, 2008.

\bibitem[Belkin and Niyogi(2003)]{belkin2003laplacian}
Mikhail Belkin and Partha Niyogi.
\newblock Laplacian eigenmaps for dimensionality reduction and data
  representation.
\newblock \emph{Neural computation}, 15\penalty0 (6):\penalty0 1373--1396,
  2003.

\bibitem[Bronstein and Kokkinos(2010)]{bronstein2010scale}
Michael~M Bronstein and Iasonas Kokkinos.
\newblock Scale-invariant heat kernel signatures for non-rigid shape
  recognition.
\newblock In \emph{Computer Vision and Pattern Recognition (CVPR), 2010 IEEE
  Conference on}, pages 1704--1711. IEEE, 2010.

\bibitem[Kuang et~al.(2015)Kuang, Li, Lv, Weiwei, and Liu]{kuang2015modal}
Zhenzhong Kuang, Zongmin Li, Qian Lv, Tian Weiwei, and Yujie Liu.
\newblock Modal function transformation for isometric 3d shape representation.
\newblock \emph{Computers \& Graphics}, 46:\penalty0 209--220, 2015.

\bibitem[Gao et~al.(2014)Gao, Yu, and Pang]{gao2014compact}
Zhanheng Gao, Zeyun Yu, and Xiaoli Pang.
\newblock A compact shape descriptor for triangular surface meshes.
\newblock \emph{Computer-Aided Design}, 53:\penalty0 62--69, 2014.

\bibitem[L{\'e}vy(2006)]{levy2006laplace}
Bruno L{\'e}vy.
\newblock Laplace-beltrami eigenfunctions towards an algorithm that"
  understands" geometry.
\newblock In \emph{Shape Modeling and Applications, 2006. SMI 2006. IEEE
  International Conference on}, pages 13--13. IEEE, 2006.

\bibitem[Taubin(1995)]{taubin1995signal}
Gabriel Taubin.
\newblock A signal processing approach to fair surface design.
\newblock In \emph{Proceedings of the 22nd annual conference on Computer
  graphics and interactive techniques}, pages 351--358. ACM, 1995.

\bibitem[Mayer(2001)]{mayer2001numerical}
Uwe~F Mayer.
\newblock Numerical solutions for the surface diffusion flow in three space
  dimensions.
\newblock \emph{Computational and Applied Mathematics}, 20\penalty0
  (3):\penalty0 361--379, 2001.

\bibitem[Xu(2004)]{xu2004discrete}
Guoliang Xu.
\newblock Discrete laplace--beltrami operators and their convergence.
\newblock \emph{Computer aided geometric design}, 21\penalty0 (8):\penalty0
  767--784, 2004.

\bibitem[Desbrun et~al.(1999)Desbrun, Meyer, Schr{\"o}der, and
  Barr]{desbrun1999implicit}
Mathieu Desbrun, Mark Meyer, Peter Schr{\"o}der, and Alan~H Barr.
\newblock Implicit fairing of irregular meshes using diffusion and curvature
  flow.
\newblock In \emph{Proceedings of the 26th annual conference on Computer
  graphics and interactive techniques}, pages 317--324. ACM
  Press/Addison-Wesley Publishing Co., 1999.

\bibitem[Meyer et~al.(2003)Meyer, Desbrun, Schr{\"o}der, and
  Barr]{meyer2003discrete}
Mark Meyer, Mathieu Desbrun, Peter Schr{\"o}der, and Alan~H Barr.
\newblock Discrete differential-geometry operators for triangulated
  2-manifolds.
\newblock In \emph{Visualization and mathematics III}, pages 35--57. Springer,
  2003.

\bibitem[Xu(2006)]{xu2006convergence}
Guoliang Xu.
\newblock Convergence analysis of a discretization scheme for gaussian
  curvature over triangular surfaces.
\newblock \emph{Computer Aided Geometric Design}, 23\penalty0 (2):\penalty0
  193--207, 2006.

\bibitem[Tenenbaum et~al.(2000)Tenenbaum, De~Silva, and
  Langford]{tenenbaum2000global}
Joshua~B Tenenbaum, Vin De~Silva, and John~C Langford.
\newblock A global geometric framework for nonlinear dimensionality reduction.
\newblock \emph{science}, 290\penalty0 (5500):\penalty0 2319--2323, 2000.

\bibitem[Silva and Tenenbaum(2003)]{silva2003global}
Vin~D Silva and Joshua~B Tenenbaum.
\newblock Global versus local methods in nonlinear dimensionality reduction.
\newblock In \emph{Advances in neural information processing systems}, pages
  721--728, 2003.

\bibitem[Roweis and Saul(2000)]{roweis2000nonlinear}
Sam~T Roweis and Lawrence~K Saul.
\newblock Nonlinear dimensionality reduction by locally linear embedding.
\newblock \emph{science}, 290\penalty0 (5500):\penalty0 2323--2326, 2000.

\bibitem[Saul and Roweis(2003)]{saul2003think}
Lawrence~K Saul and Sam~T Roweis.
\newblock Think globally, fit locally: unsupervised learning of low dimensional
  manifolds.
\newblock \emph{Journal of machine learning research}, 4\penalty0
  (Jun):\penalty0 119--155, 2003.

\bibitem[Coifman and Lafon(2006)]{coifman2006diffusion}
Ronald~R Coifman and St{\'e}phane Lafon.
\newblock Diffusion maps.
\newblock \emph{Applied and computational harmonic analysis}, 21\penalty0
  (1):\penalty0 5--30, 2006.

\bibitem[Belkin and Niyogi(2002)]{belkin2002laplacian}
Mikhail Belkin and Partha Niyogi.
\newblock Laplacian eigenmaps and spectral techniques for embedding and
  clustering.
\newblock In \emph{Advances in neural information processing systems (NIPS)},
  pages 585--591, 2002.

\bibitem[Wachinger and Navab(2010{\natexlab{a}})]{wachinger2010manifold}
Christian Wachinger and Nassir Navab.
\newblock Manifold learning for multi-modal image registration.
\newblock In \emph{BMVC}, pages 1--12. Citeseer, 2010{\natexlab{a}}.

\bibitem[Wachinger and Navab(2010{\natexlab{b}})]{wachinger2010structural}
Christian Wachinger and Nassir Navab.
\newblock Structural image representation for image registration.
\newblock In \emph{Computer Vision and Pattern Recognition Workshops (CVPRW),
  2010 IEEE Computer Society Conference on}, pages 23--30. IEEE,
  2010{\natexlab{b}}.

\bibitem[Dulmage and Mendelsohn(1958)]{dulmage1958coverings}
Andrew~L Dulmage and Nathan~S Mendelsohn.
\newblock Coverings of bipartite graphs.
\newblock \emph{Canadian Journal of Mathematics}, 10\penalty0 (4):\penalty0
  516--534, 1958.

\bibitem[Weyl(1911)]{weyl1911asymptotische}
Hermann Weyl.
\newblock {\"U}ber die asymptotische verteilung der eigenwerte.
\newblock \emph{Nachrichten von der Gesellschaft der Wissenschaften zu
  G{\"o}ttingen, Mathematisch-Physikalische Klasse}, 1911:\penalty0 110--117,
  1911.

\bibitem[Bronstein et~al.(2008)Bronstein, Bronstein, and
  Kimmel]{bronstein2008numerical}
Alexander~M Bronstein, Michael~M Bronstein, and Ron Kimmel.
\newblock \emph{Numerical geometry of non-rigid shapes}.
\newblock Springer Science \& Business Media, 2008.

\bibitem[Siddiqi et~al.(2008)Siddiqi, Zhang, Macrini, Shokoufandeh, Bouix, and
  Dickinson]{siddiqi2008retrieving}
Kaleem Siddiqi, Juan Zhang, Diego Macrini, Ali Shokoufandeh, Sylvain Bouix, and
  Sven Dickinson.
\newblock Retrieving articulated 3-d models using medial surfaces.
\newblock \emph{Machine vision and applications}, 19\penalty0 (4):\penalty0
  261--275, 2008.

\bibitem[Mirloo and Ebrahimnezhad(2017)]{Mirloo2017}
Mahsa Mirloo and Hossein Ebrahimnezhad.
\newblock Non-rigid 3d object retrieval using directional graph representation
  of wave kernel signature.
\newblock \emph{Multimedia Tools and Applications}, Apr 2017.
\newblock ISSN 1573-7721.
\newblock \doi{10.1007/s11042-017-4617-x}.
\newblock URL \url{https://doi.org/10.1007/s11042-017-4617-x}.

\bibitem[Masoumi and Hamza(2017)]{masoumi2017global}
Majid Masoumi and A~Ben Hamza.
\newblock Global spectral graph wavelet signature for surface analysis of
  carpal bones.
\newblock \emph{arXiv preprint arXiv:1709.02782}, 2017.

\bibitem[Boscaini et~al.(2015)Boscaini, Masci, Melzi, Bronstein, Castellani,
  and Vandergheynst]{boscaini2015learning}
Davide Boscaini, Jonathan Masci, Simone Melzi, Michael~M Bronstein, Umberto
  Castellani, and Pierre Vandergheynst.
\newblock Learning class-specific descriptors for deformable shapes using
  localized spectral convolutional networks.
\newblock \emph{Computer Graphics Forum}, 34\penalty0 (5):\penalty0 13--23,
  2015.

\bibitem[Li and Hamza(2014)]{li2014spatially}
Chunyuan Li and A~Ben Hamza.
\newblock Spatially aggregating spectral descriptors for nonrigid 3d shape
  retrieval: a comparative survey.
\newblock \emph{Multimedia Systems}, 20\penalty0 (3):\penalty0 253--281, 2014.

\bibitem[Lian et~al.(2011)Lian, Godil, Bustos, Daoudi, Hermans, Kawamura,
  Kurita, Lavoua, and Dp~Suetens]{lian2011shape}
Z~Lian, A~Godil, B~Bustos, M~Daoudi, J~Hermans, S~Kawamura, Y~Kurita, G~Lavoua,
  and P~Dp~Suetens.
\newblock Shrec’11 track: Shape retrieval on non-rigid 3d watertight meshes.
\newblock In \emph{Eurographics Workshop on 3D Object Retrieval (3DOR)}, 2011.

\bibitem[Shilane et~al.(2004)Shilane, Min, Kazhdan, and
  Funkhouser]{shilane2004princeton}
Philip Shilane, Patrick Min, Michael Kazhdan, and Thomas Funkhouser.
\newblock The princeton shape benchmark.
\newblock In \emph{Shape modeling applications, 2004. Proceedings}, pages
  167--178. IEEE, 2004.

\bibitem[Liu et~al.(2011)Liu, Chen, and Tang]{liu2011construction}
Yong-Jin Liu, Zhanqing Chen, and Kai Tang.
\newblock Construction of iso-contours, bisectors, and voronoi diagrams on
  triangulated surfaces.
\newblock \emph{IEEE Transactions on Pattern Analysis and Machine
  Intelligence}, 33\penalty0 (8):\penalty0 1502--1517, 2011.

\bibitem[Zhang et~al.(2009)Zhang, Bajaj, and Xu]{zhang2009surface}
Yongjie Zhang, Chandrajit Bajaj, and Guoliang Xu.
\newblock Surface smoothing and quality improvement of quadrilateral/hexahedral
  meshes with geometric flow.
\newblock \emph{International Journal for Numerical Methods in Biomedical
  Engineering}, 25\penalty0 (1):\penalty0 1--18, 2009.

\end{thebibliography}

\end{document}